\documentclass[preprint2,numberedappendix]{emulateapj-rtx4}
\usepackage{graphicx,graphics,amsmath}
\usepackage{graphicx,natbib,color}
\usepackage{bm,url}
\graphicspath{{.}} %for arxive
\def\blue{\textcolor{black}} %for arxive
\newcommand{\EQ}{\begin{equation}}
\newcommand{\EN}{\end{equation}}
\newcommand{\EQA}{\begin{eqnarray}}
\newcommand{\ENA}{\end{eqnarray}}
\newcommand{\nab}{\mbox{\boldmath $\nabla$} {}}

\newcommand{\ghat}{\hat{g}}

\newcommand{\AAA}{\bm{{A}}}
\newcommand{\BB}{\bm{{B}}}
\newcommand{\JJ}{\bm{{J}}}
\newcommand{\UU}{\bm{{U}}}

\newcommand{\meanB}{\overline{B}}

\newcommand{\meanBB}{\bm{\overline{{B}}}}

\newcommand{\ppi}{\bm{{{\pi}}}}

\newcommand{\bb}{\bm{{{b}}}}

\newcommand{\ff}{\bm{{{f}}}}
\newcommand{\kk}{\bm{{{k}}}}
\newcommand{\xx}{\bm{{{x}}}}

\newcommand{\eee}{\hat{\mbox{\boldmath $e$}} {}}

\def\Gr{\mbox{\rm Gr}}

\def\kf{k_\mathrm{f}}
\def\ii{{\rm i}}
\def\dd{{\rm d}}

\def\Pm{{\rm Pr}_\mathrm{M}}
\def\Rm{{\rm Re}_\mathrm{M}}

\def\Rey{\mbox{\rm Re}}

\def\rstar{r_\star}
\def\tautd{\tau_{\rm td}}
\def\Peff{{\cal P}_{\rm eff}}
\newcommand{\meanrho}{\overline{\rho}}
\def\qp{q_{\rm p}}
\def\qs{q_{\rm s}}
\def\qg{q_{\rm g}}
\def \fftilde {\tilde{{\bm f}}}
\def \smax {\sigma_{\rm max}}
\def \wf {w_{\rm f}}
\newcommand{\erf}{{\rm erf}}
\newcommand{\DD}{{\rm D} {}}
\newcommand{\grav}{\mbox{\boldmath $g$} {}}
\newcommand{\etal}{et al.}

\newcommand{\betap}{\beta_{\rm p}}

\newcommand{\etatz}{\eta_{\rm t0}}

\newcommand{\cs}{c_{\rm s}}
\newcommand{\urms}{u_{\rm rms}}

\newcommand{\Beq}{B_{\rm eq}}
\newcommand{\Beqz}{B_{\rm eq0}}

\def\onethird{{\textstyle{1\over3}}}
\def\half{{\textstyle{1\over2}}}

\def\onethird{{\textstyle{1\over3}}}

\newcommand{\RRRR}{\mbox{\boldmath ${\sf R}$} {}}
\newcommand{\SSSS}{\mbox{\boldmath ${\sf S}$} {}}

\newcommand{\Eq}[1]{Equation~(\ref{#1})}

\newcommand{\Sec}[1]{\S\ref{#1}}

\newcommand{\Fig}[1]{Figure~\ref{#1}}
\newcommand{\Tab}[1]{Table~\ref{#1}}
\newcommand{\Figs}[2]{Figures~\ref{#1} and \ref{#2}}

\newcommand{\bra}[1]{\langle #1\rangle}
\begin{document}

\title{Bipolar magnetic spots from dynamos in stratified spherical shell turbulence}
\author{Sarah Jabbari$^{1,2}$, Axel Brandenburg$^{1,2}$,
Nathan Kleeorin$^{3,1}$, Dhrubaditya Mitra$^1$,
Igor Rogachevskii$^{3,1}$}
\affil{$^1$Nordita, KTH Royal Institute of Technology and Stockholm University,
Roslagstullsbacken 23, SE-10691 Stockholm, Sweden\\
$^2$Department of Astronomy, AlbaNova University Center, Stockholm University, SE-10691 Stockholm, Sweden\\
$^3$Department of Mechanical Engineering, Ben-Gurion University of the Negev,
POB 653, Beer-Sheva 84105, Israel
}

\email{sarahjab@kth.se
($ $Revision: 1.2 $ $)
}

\begin{abstract}
Recent work by Mitra et al.\ (2014) has shown that in strongly stratified
forced two-layer turbulence with helicity and corresponding large-scale dynamo
action in the lower layer, and nonhelical turbulence in the upper,
a magnetic field occurs in the upper layer
in the form of sharply bounded bipolar magnetic spots.
Here we extend this model to spherical wedge geometry covering the
northern hemisphere up to $75^\circ$ latitude and an azimuthal extent
of $180^\circ$.
The kinetic helicity and therefore also the large-scale magnetic field
are strongest at low latitudes.
For moderately strong stratification, several bipolar spots form that
fill eventually the full longitudinal extent.
At early times, the polarity of spots reflects the orientation of the
underlying azimuthal field, as expected from Parker's $\Omega$-shaped flux loops.
At late times their tilt changes such that there is a radial field of
opposite orientation at different latitudes separated by about $10^\circ$.
\blue{Our model demonstrates} the spontaneous formation of
spots of sizes much larger than the pressure scale height.
Their tendency to produce filling factors close to unity is argued to be
reminiscent of highly active stars.
We confirm that strong stratification
and strong scale separation are
essential ingredients behind magnetic spot formation,
which appears to be associated with downflows at larger depths.
\end{abstract}

\section{Introduction}

Solar activity is characterized by the formation of
magnetic spots.
Sunspots are relatively small concentrations of magnetic field at the
surface, where the radiation is suppressed significantly, making these
regions cooler than their surroundings.
Similar phenomena are also expected to occur on other stars exhibiting
magnetic cycles, although the starspots observed so far all tend to
be significantly larger than sunspots and not necessarily spatially
correlated with the surface temperature \citep{IIPeg13}.
There is little doubt that magnetic spots are associated with an
underlying dynamo in the outer convection zones of these stars, but
it is not clear whether they are caused by deeply rooted magnetic flux tubes at
the bottom of the convection zones \citep{Caligari}, or whether they are
merely shallow magnetic concentrations formed locally where the
near-surface magnetic field exceeds a certain threshold.
Magnetic field visualizations from convectively driven dynamo simulations
have shown serpentine-shaped flux tubes that can be expected to
intersect the surface \blue{\citep{NM14,FF14}}, but these tubes would expand during
their ascent, so some sort of re-amplification of these tubes would be
needed to explain sunspots.

The foundations of magnetic spot formation have been developed by
\cite{Par55}, who identified magnetic buoyancy as the main agent
responsible for bringing magnetic fields to the surface.
In his subsequent work, \cite{Par79} identified the need for a special
mechanism to make these concentrations sufficiently cool and evacuated
so as to explain the observed values of the magnetic field in sunspots.
He postulated the existence of suitable downflows that would help
to evacuate the magnetic flux tube in its upper parts.

The appearance of such downdrafts in the region of spontaneously
formed magnetic spots has been observed in the numerical simulations
of \cite{BKR13} using forced turbulence with weak imposed
vertical magnetic field.
Downdrafts have also been seen in simulations of buoyantly rising
flux tubes some time after they reached the surface \citep{RC14}.
A possible mechanism for producing such downflows might well be
the negative effective magnetic pressure instability (NEMPI).
It is based on the magnetic suppression of
the total (hydrodynamic plus magnetic)
turbulent pressure.
The importance of the difference in turbulent pressure inside and
outside magnetic structures was first emphasized by \cite{vanBall84}.
Subsequent mean-field calculations have shown that, if the magnetic Reynolds
number is larger than unity, the effective
large-scale magnetic pressure (sum of turbulent and non-turbulent
contributions) becomes negative and a large-scale instability (namely NEMPI)
can be excited \citep{KRR89,KRR90,KMR96,KR94,RK07}.
This instability redistributes magnetic flux and
can cause the formation of magnetic structures.
As the work of recent years has shown,
for horizontal magnetic fields, negative effective magnetic pressure leads
to negative magnetic buoyancy at sub-equipartition field strengths
\citep{BKR10,BKKMR11}.
However, for vertical fields the return flow replenishing the downflow
occurs predominantly along magnetic field lines and has a much
larger impact near the surface, where it can lead to super-equipartition
strength flux concentrations \citep{BKR13}.
Corresponding mean-field simulations \citep{BGJKR14} have displayed
great similarity with Parker's original picture \citep{Par79}, where
he explicitly stated the need for postulating the existence of downdrafts,
leaving however the question about their origin open.
On the other hand, the downflows have a strengths of only about 20\%
of the turbulent rms velocity and are therefore not easily recognized
among the downflows due to convection.
Depending on circumstances, NEMPI can also lead to the
formation of bipolar spots
\citep{WLBKR13,WLBKR15} with super-equipartition field strengths \citep{Mit14}.
The latter looks remarkably similar to the bipolar regions found by
\cite{SN12} in realistic simulations of solar convection with an
unstructured magnetic field being supplied at the bottom of their domain.

Strong stratification is a prerequisite for NEMPI to occur.
In recent direct numerical simulations (DNS) of \cite{Mit14}, it was
demonstrated that stratification plays a crucial role in the formation of
magnetic spots that have surprisingly sharp boundaries
with opposite polarities on the two sides.
In these simulations, the turbulence in the deeper parts was made
helical such that a large-scale magnetic field can be generated
by the $\alpha$ effect associated with the kinetic helicity of the turbulence.
In the upper part of the domain in the DNS of \cite{Mit14} the
forcing was non-helical, so there is no $\alpha$ effect,
but NEMPI can still work.
What is surprising in those simulations is the fact that
the magnetic fields in these structures exceeds the equipartition value
by a factor of three or more.
Since the effective magnetic pressure has not been measured
in the simulations of \cite{Mit14}, it is not obvious
that the underlying mechanism is actually related to NEMPI,
even though downflows with a strength of about 20\% of the
turbulent velocity have been detected by \cite{Mit14}.
The physics of the formation of sharp magnetic edges of
bipolar structures in the turbulent flow remains however elusive.

The magnetic field from the dynamo simulations of \cite{Mit14} had the
property of extending over the entire horizontal length of the domain.
As a consequence, only one bipolar structure was produced, which
occasionally developed horizontal bands extending again over the full
length of the horizontally periodic domain.
To overcome this artifact of periodic boundary conditions, it is important
to consider larger domains with either no or at least with physically motivated
boundary conditions.
A spherical shell is an obvious choice.

The dynamics of NEMPI from dynamo-generated magnetic fields in spherical
shells or wedges has recently been studied by \cite{Jab13}
in a mean-field simulations (MFS),
although in their case the dynamo mechanism operated throughout the
domain and not just below a certain depth, as in the DNS of \cite{Mit14}.
Here we combine the two-layer setup of \cite{Mit14} with the
shell geometry used in the MFS of \cite{Jab13}.
There is also another DNS study by \cite{Jab14} in which they investigated
a combined system of dynamo and NEMPI in
Cartesian geometry.
In that paper, the combination of rotation and stratification
leads to an $\alpha^2$ dynamo.
In the present work, we ignore rotation to understand first a simpler case
using instead helically forced turbulence.
Investigation of a similar system with rotation and shear
will be the subject of a future study.

The purpose of the present paper is to study forced turbulence in a
strongly stratified spherical shell.
As in \cite{Mit14}, the turbulence is made helical below a certain
radius $\rstar$ so as to enable the formation of large-scale magnetic fields
by the $\alpha^2$ dynamo mechanism.
The sign of the helicity is assumed to change across the equator.
This leads to the formation of dynamo waves that travel toward the
equator with opposite polarity in the southern hemisphere \citep{Mit10}.
This is a property that has been associated with the choice of
perfectly conducting boundary conditions at high latitudes of the wedge.
On the other hand, changing the high-latitude boundary condition to
a normal field condition
(often referred to as a vertical field condition)
causes dynamo waves to propagate away from the
equator and toward high latitudes, but now with the same polarity in
both hemispheres \citep{BCC09}.
In the present paper, we reconsider the former case and apply a suitable
boundary condition at the equator to cut the computational costs.

There is another potential artifact of the dynamo waves of \cite{Mit10}
in that they tend to occur at high latitudes where the kinetic helicity
is assumed largest.
On the other hand, we have seen in DNS of turbulent convection in spherical
shells that kinetic helicity is in fact concentrated to regions
outside the inner tangent cylinder of the shell \citep{KMB12}.
This restricts the kinetic helicity essentially to low latitudes below
$45\degr$.
We model this feature here by assuming a suitable profile for the kinetic
helicity of the forcing function in the deeper parts of the shell.

\section{The model}

In this paper we investigate a system similar to that of \cite{Mit14}, but in
spherical geometry assuming either symmetric (quadrupolar)
or anti-symmetric (dipolar) field properties about the equator.
\cite{Jab13} used MFS in spherical geometry
to show how the large-scale dynamo can interact
with NEMPI in such a coupled system.
In their MFS, the mean-field Lorentz force was parameterized,
which is subject to uncertainties.
It is therefore useful to perform DNS and to study how the results depend
on domain size, density stratification, geometry, and
boundary conditions.
As explained above, the
main difference here is the fact that the forcing is not uniform
in whole domain.
As in \cite{Mit14}, our domain is divided into two parts.
We apply helical forcing (which leads to an alpha squared dynamo)
in the lower part of the domain and non-helical forcing in
the upper part of the domain.
The position of the border between these two areas is varied
to see how it affects the results.
We expect to detect similar intense bipolar region of
earlier DNS of \cite{Mit14}.

We use an isothermal equation of state, so no convection is possible.
Therefore, turbulence is driven using volume forcing given by a function $f$
that is $\delta$-correlated in time and monochromatic in space.
It consists of random non-polarized waves whose direction
and phase change randomly at each time step.
We present a more detailed discussion about forcing in \Sec{forcing}.

\subsection{Basic equations}

In DNS of an isothermally stratified layer we solve the equations
for the velocity $\UU$, the magnetic vector potential $\AAA$, and
the density $\rho$,
\begin{equation}
\rho{\DD\UU\over\DD t}=\JJ\times\BB-\cs^2\nab\rho+\nab\cdot(2\nu\rho\SSSS)
+\rho(\ff+\grav),
\end{equation}
\begin{equation}
{\partial\AAA\over\partial t}=\UU\times\BB+\eta\nabla^2\AAA,
\end{equation}
\begin{equation}
{\partial\rho\over\partial t}=-\nab\cdot\rho\UU,
\end{equation}
where the operator $\DD/\DD t=\partial/\partial t+\UU\cdot\nab$ is
the advective derivative, $\eta$ is the magnetic diffusivity,
$\BB=\nab\times\AAA$ is the magnetic field,
$\JJ=\nab\times\BB/\mu_0$ is the current density,
${\sf S}_{ij}=\half(U_{i,j}+U_{j,i})-\onethird\delta_{ij}\nab\cdot\UU$
is the traceless rate of strain tensor (the commas denote
partial differentiation), $\nu$ is the kinematic viscosity,
$\cs$ is the isothermal sound speed, and $\mu_0$ is the vacuum permeability.
We adopt spherical coordinates ($r, \theta, \phi$).

For the $\phi$ direction we use periodic boundary conditions.
In the radial direction (the direction of the stratification), we
consider perfectly conducting boundary conditions at the bottom
and a vertical field boundary condition at the top.
At the equator, we adopt a symmetry condition such that the magnetic
field is either symmetric (quadrupolar) or anti-symmetric (dipolar)
with respect to the equator.
For the magnetic field we assume perfect conductor boundary conditions
on the latitudinal ($\theta=\theta_0$) and lower radial ($r=r_0$) boundaries,
and radial field boundary conditions on the outer radius ($r=R$).
On the equator, we assume either dipolar or a quadrupolar symmetry.
In terms of the magnetic vector potential these conditions translate to
\begin{equation}
\frac{\partial A_r}{\partial r}= A_\theta=A_\phi =0 \,\quad
(r=r_0),
\end{equation}
\begin{equation}
A_r=0, \;\; \frac{\partial A_{\theta}}{\partial r}=-\frac{A_{\theta}}{r},\;\; \frac{\partial
A_{\phi}}{\partial r}=-\frac{A_{\phi}}{r} \quad (r=R),
\end{equation}
\begin{equation}
A_r=\frac{\partial A_\theta}{\partial\theta}=A_\phi=0 \quad
(\theta=\theta_0,\pi/2)
\end{equation}
for quadrupolar symmetry and
\begin{eqnarray}
\frac{\partial A_r}{\partial\theta}=A_\theta=
\frac{\partial A_\phi}{\partial\theta}=0 \quad
(\theta=\pi/2)
\end{eqnarray}
for dipolar symmetry.

For the velocity field we use stress-free,
non-penetrating boundary conditions in the radial direction.
The gravitational acceleration is $\grav=-\nab\Phi$, where
\EQ
\Phi(r)=-GM\left({1\over r}-{1\over r_{\rm m}}\right).
\label{Phir}
\EN
Here $G$ is Newton's constant and $M$ is the mass of the sphere (or star).
For an isothermal gas, the hydrostatic density stratification obeys
$\rho=\rho_0\exp(-\Phi/\cs^2)$, where $\rho=\rho_0$ is the density
in the middle of the shell at $r=r_{\rm m}=(r_0+R)/2$.
The radial component of the gravitational acceleration is then $g=-GM/r^2$.
The quantity $GM$ determines the density contrast
$\Gamma_\rho=\rho_{\rm bot}/\rho_{\rm top}$ between bottom and top
of the domain.
\blue{
Initially, we have $\Gamma_\rho=\exp(R/r_0-1)^{GM/R\cs^2}$.
The density scale height is given by $H_\rho=\cs^2/GM$.
}
The thickness of the shell is $\Delta r=R-r_0$, and it is used to define
a reference wavenumber $k_1=2\pi/\Delta r$.

\subsection{The forcing function}
\label{forcing}

The forcing function $\ff$ is similar to that of \cite{Mit14},
\begin{equation}
\ff(\xx,t)={\rm Re}\left[N\fftilde(\kk,t)\exp(\ii\kk\cdot\xx+\ii\varphi)\right],
\end{equation}
where $\xx$ is the position vector,
$-\pi<\varphi\le\pi$ is a randomly selected phase,
and $\kk$ is the wavevector which is chosen from a set of
wavevectors in a certain range around a given forcing wavenumber, $\kf$.
The Fourier amplitudes, $\fftilde({\kk})$, are defined as
\begin{equation}
\fftilde({\kk})=\RRRR\cdot\fftilde({\kk})^{\rm(nohel)}\quad\mbox{with}\quad
{\sf R}_{ij}={\delta_{ij}-\ii\sigma\epsilon_{ijk}\hat{k}
\over\sqrt{1+\sigma^2}},
\end{equation}
where $\sigma$ characterizes the fractional helicity of $\ff$, and
\begin{equation}
\fftilde({\kk})^{\rm(nohel)}=
\left(\kk\times\eee\right)\left/\sqrt{\kk^2-(\kk\cdot\eee)^2}\right.
\label{nohel_forcing}
\end{equation}
is a non-helical forcing function, and $\eee$ is an arbitrary unit vector
not aligned with $\kk$ and $\hat{\kk}$ is the unit vector along $\kk$;
note that $|\fftilde|^2=1$.
The degree of helicity is modulated in space via the function
\begin{equation}
\sigma(r,\theta) = \frac{\smax}{2}
\left[1-\erf\left(\frac{r-r_\ast}{\wf}\right) \right]
\cos\theta \sin^n\!\theta,
\label{alpha_profile}
\end{equation}
where $\erf$ is the error function,
$r_\ast$ is the radius above which the helicity vanishes,
$\wf$ is the width of the transition layer,
and the exponent $n$ determines the latitudinal helicity profile.
We choose $\wf=0.01$ for all the simulations.
\blue{
The amplitude of the forcing is, however, independent of $r$ and therefore
also the root-mean-square velocity is essentially independent of $r$.
}
For more details of this type of forcing see \cite{Mit14}.

\blue{
We note that the degree of helicity of the forcing function
is here assumed to be independent of the degree of stratification.
In reality, of cause, helicity is actually a consequence of stratification
together with rotation \citep{KR80}.
We return to this question in the conclusions, where we discuss
possible artifacts resulting from this assumption.
}

\subsection{Parameters of the simulations}

During the exponential growth phase of the dynamo, the growth rate
is calculated as $\lambda=d\ln B_{\rm rms}/dt$.
The nondimensional growth rate is given as $\tilde{\lambda}=\lambda/\urms\kf$.
However, the time of the simulation is normally specified in terms of the
turbulent-diffusive time $\tautd=(\etatz k_1^2)^{-1}$,
where $\etatz=\urms/3\kf$ is the estimated turbulent diffusivity.
In most of the calculations, we use a scale separation ratio $\kf/k_1$ of 30
and a fluid Reynolds number $\Rey\equiv\urms/\nu\kf$ of 20.
Our magnetic Prandtl number $\Pm=\nu/\eta$ is 1, so
the magnetic Reynolds number is then $\Rm=\Pm\Rey=20$.
These values are chosen to have both $\kf$ and $\Rey$
large enough for NEMPI to develop at an affordable numerical resolution.
The magnetic field is expressed in units of local equipartition
magnetic field, $\Beq(r)=\sqrt{\mu_0\meanrho(r)}\urms$, where $\meanrho(r)$
is the density averaged over time and spherical shells.
We also define $\Beqz=\sqrt{\mu_0\rho_0} \, \urms$.
In the following, we use non-dimensional units by setting $\cs=\mu_0=\rho_0=1$.

\blue{
We perform simulations with values of $GM/R\cs^2$ between 1 and 17.
With $\exp(R/r_0-1)\approx1.54$, this implies that
$\Gamma_\rho\approx1.54^{GM/R\cs^2}$ between 1.5 and 1460
for the initial values.
In the following, however, we quote the values from the relaxed run.
We perform simulations with different values of $\Gamma_\rho$,
which enables us to study the effect of stratification on the
formation of magnetic structures.
The corresponding stratification parameter of \cite{Jab14},
$\Gr=(\kf H_\rho)^{-1}$, varies then between 0.002 (for $GM/R\cs^2=1$)
and 0.03 (for $GM/R\cs^2=17$).
Even the latter value is still rather small compared with the
value of 0.16 expected from solar mixing length theory.
Increasing the value of $\Gr$ leads to a slight decrease of the growth
rate of NEMPI compared with the theoretically expected value;
see \cite{Jab14} for details.
}

For most of the simulations, we choose $n=6$ in \Eq{alpha_profile},
i.e., the helicity is maximum at lower latitudes.
This is also the case for our two reference runs, which have $\Gamma_\rho=450$
and either quadrupolar or dipolar parity.
However, for comparison we also present cases where $n=0$.

We use the {\sc Pencil Code}\footnote{\url{http://pencil-code.googlecode.com}}
to perform direct numerical simulations.
This code uses sixth-order explicit finite differences in space
and a third-order accurate time-stepping method.
We use $r_0=0.7\,R$ and $\theta_0=15\degr$.
For runs with a $\phi$ extend of $\pi$, we use a numerical resolution of
$256\times1152\times1152$ mesh points in the $r$, $\theta$, and $\phi$
directions, and $256\times1152\times288$ for all other runs.
\Tab{Tab1} shows all runs with their parameters.

\begin{figure}[t]
\centering
%\hspace*{-0.5mm}\includegraphics[width=1.0\columnwidth]{pxy_slice_256dyn_sph_sin_g14_vbc_pi_0.75.ps}
%\hspace*{-0.5mm}\includegraphics[width=1.0\columnwidth]{pxy_slice_256dyn_sph_sin_g14_vbc_pi_0.80.ps}
%\hspace*{-0.5mm}\includegraphics[width=1.0\columnwidth]{pxy_slice_256dyn_sph_sin_g14_vbc_pi_0.95.ps}
\hspace*{-0.5mm}\includegraphics[width=1.0\columnwidth]{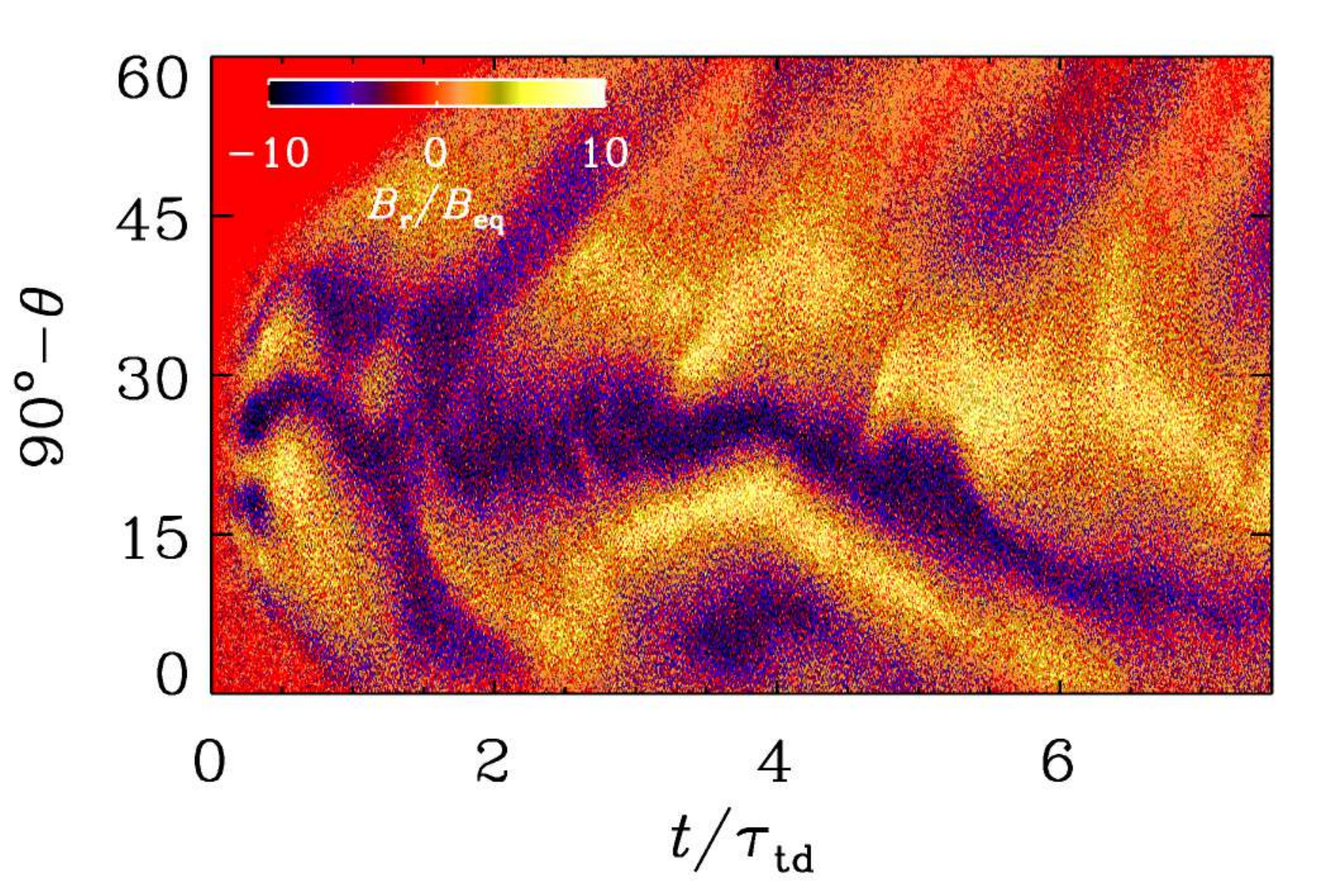}
\hspace*{-0.5mm}\includegraphics[width=1.0\columnwidth]{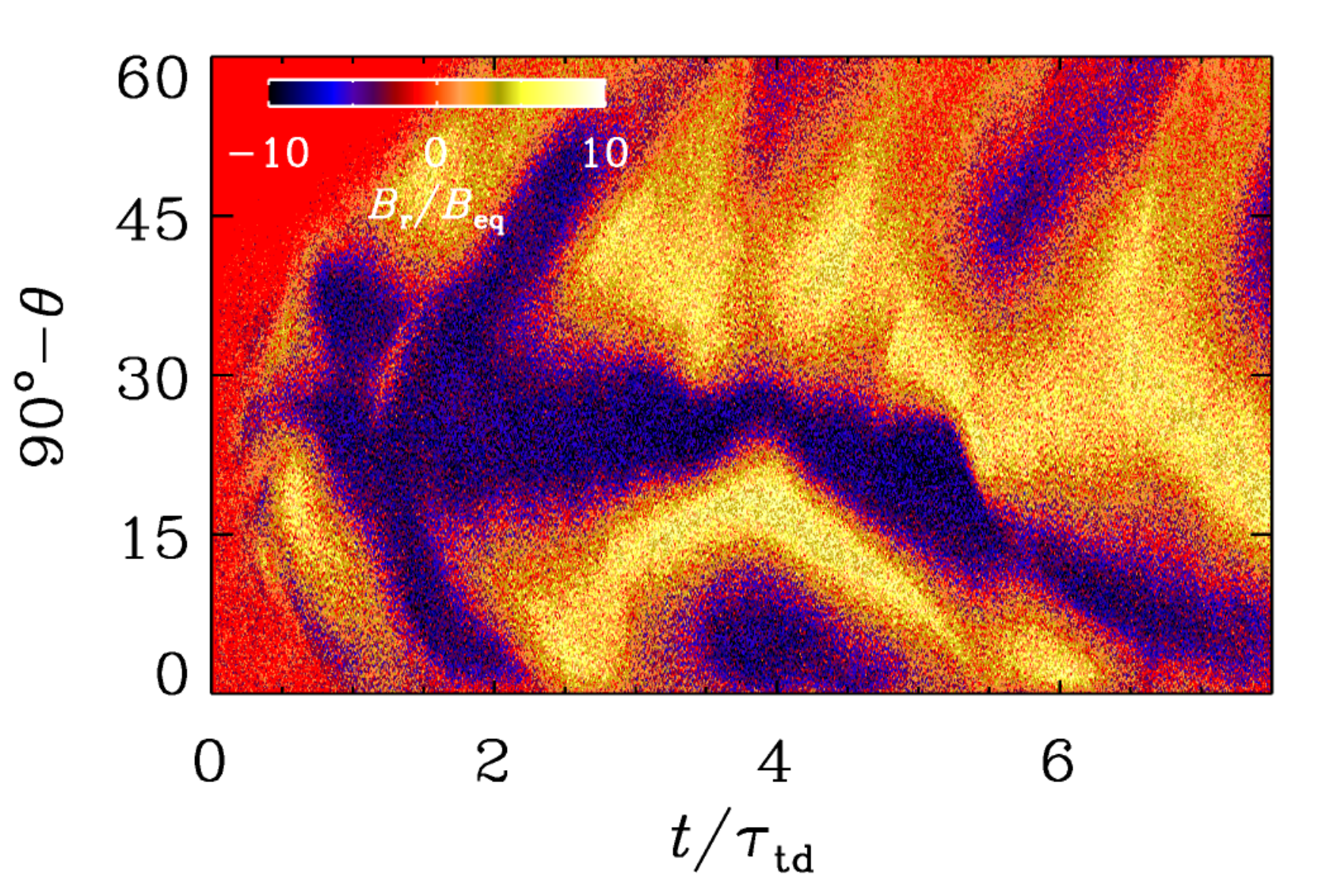}
\hspace*{-0.5mm}\includegraphics[width=1.0\columnwidth]{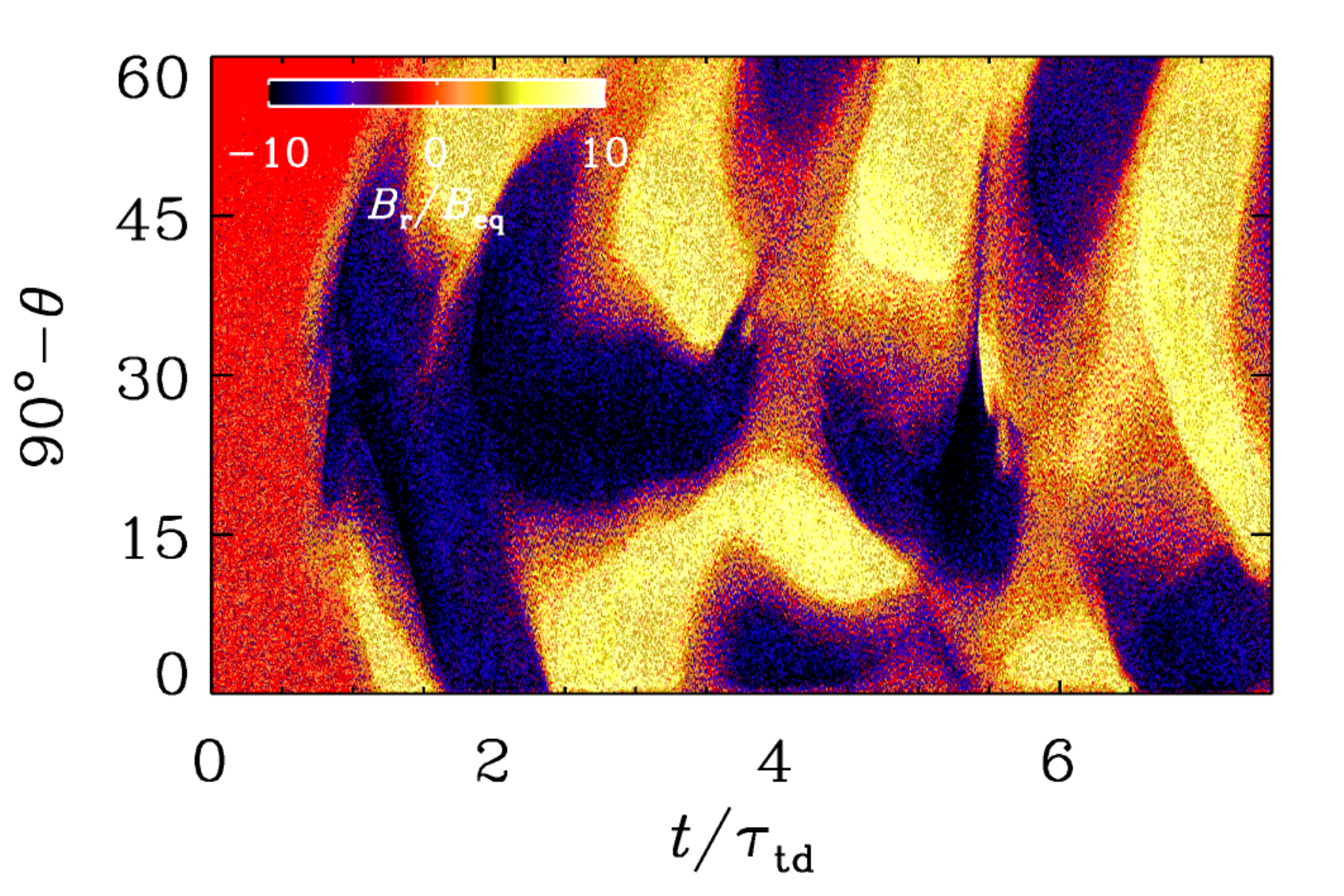}
\caption{
Butterfly diagram for our reference run (Run~\blue{D2}
with $\Gamma_\rho=450$) at
$r/R=0.75$ (top), $0.8$ (middle), and $0.95$ (bottom).
}\label{fig:butt}
\end{figure}

\begin{figure}[t]
\includegraphics[width=0.49\columnwidth]{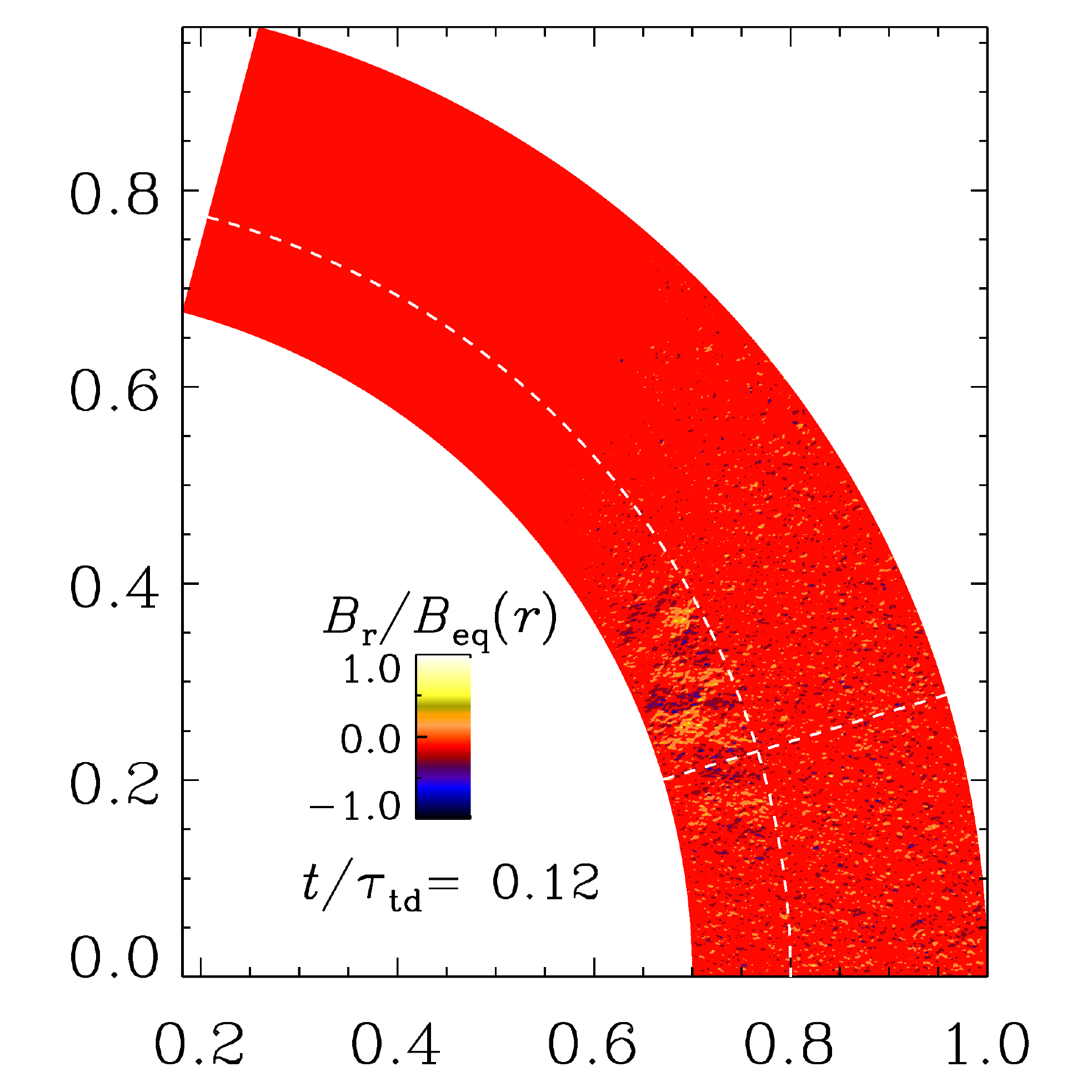}
\includegraphics[width=0.49\columnwidth]{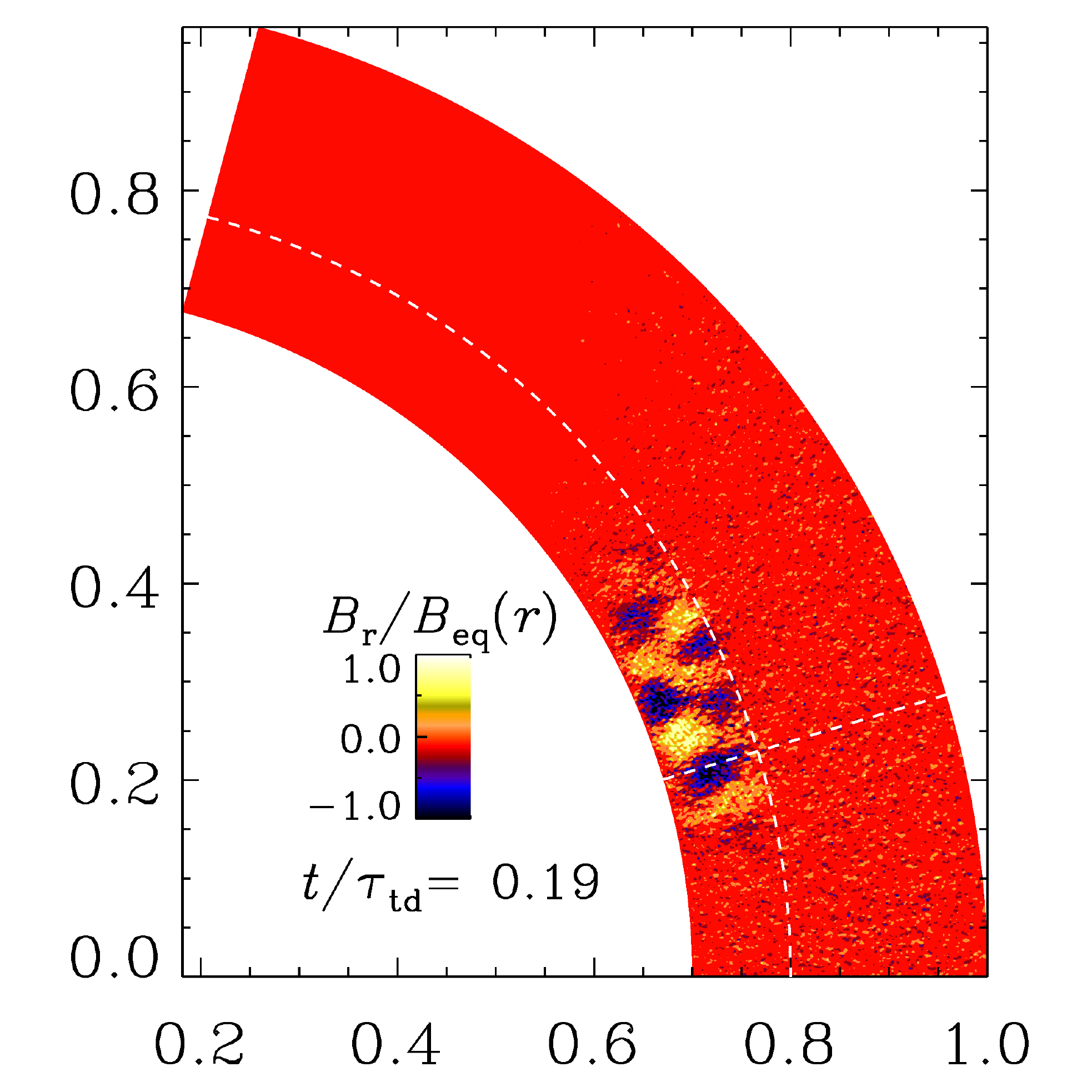}\\
\includegraphics[width=0.49\columnwidth]{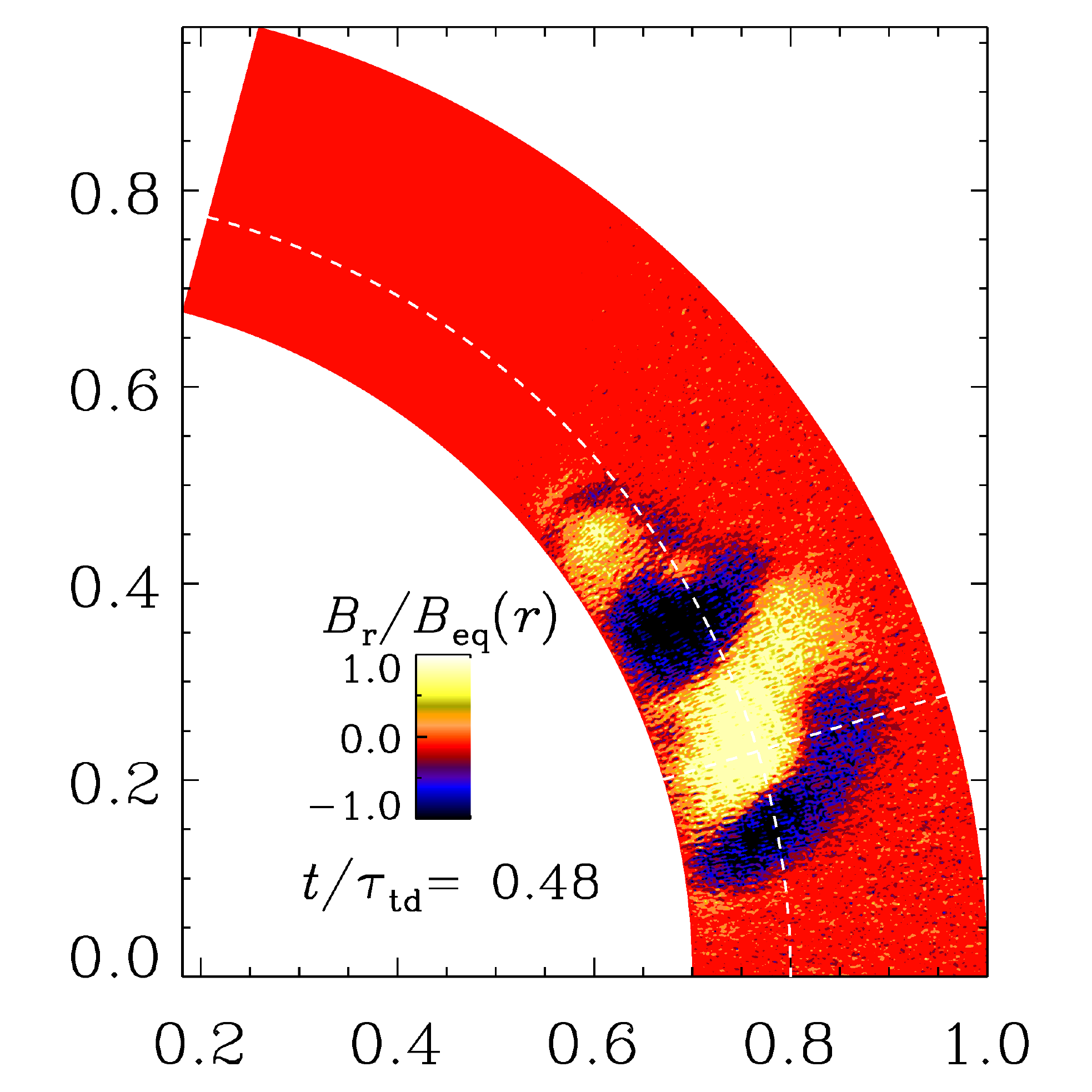}
\includegraphics[width=0.49\columnwidth]{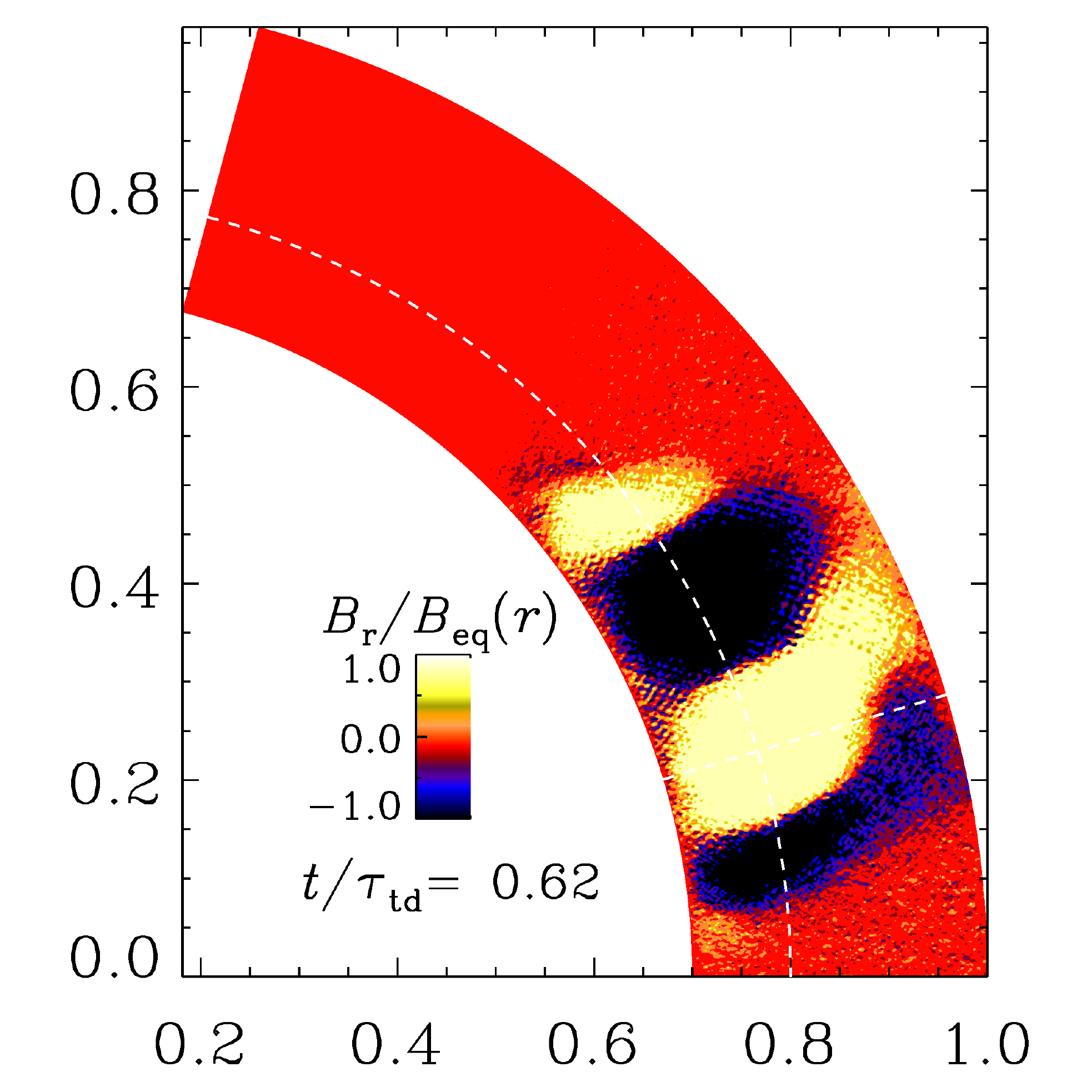}\\
\includegraphics[width=0.49\columnwidth]{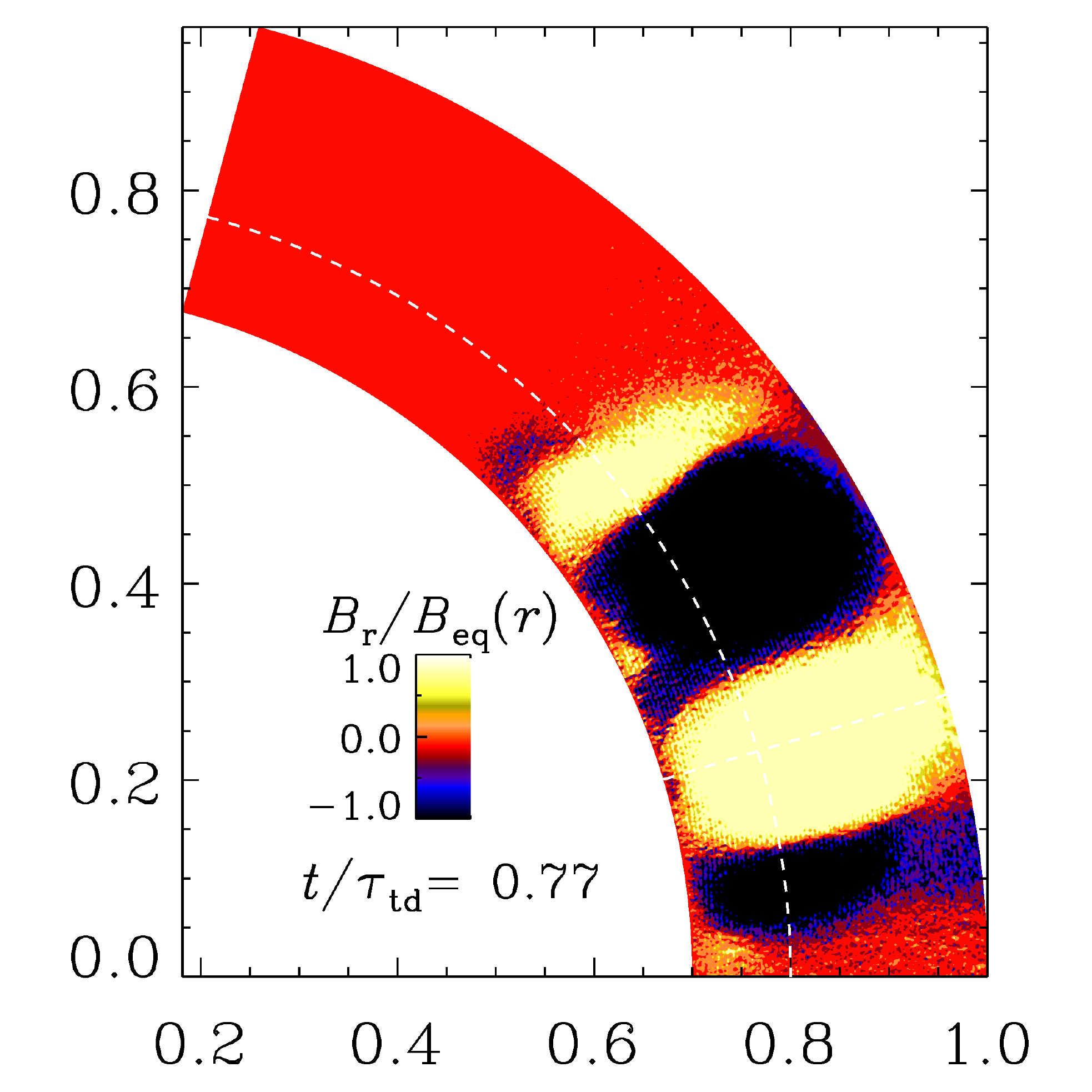}
\includegraphics[width=0.49\columnwidth]{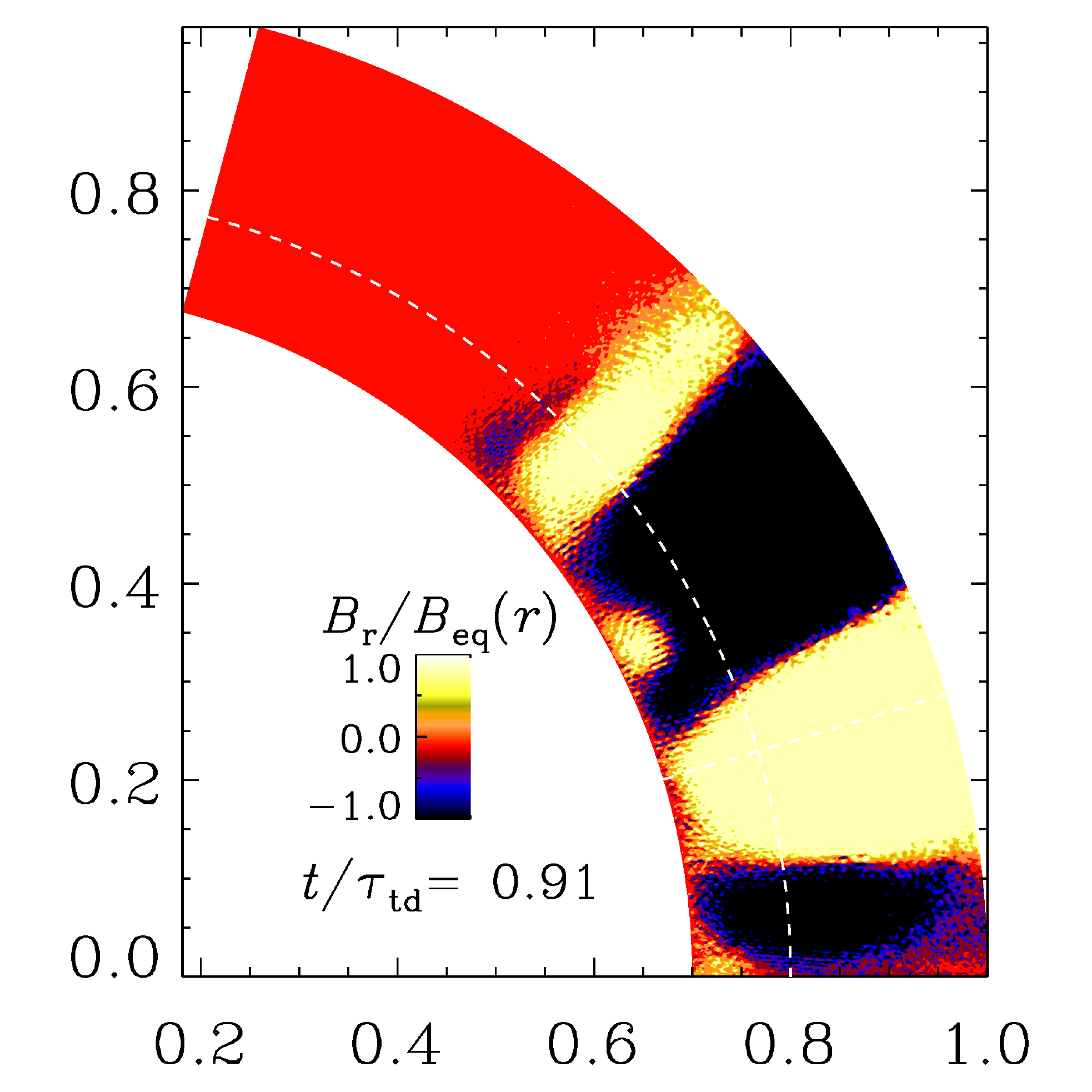}
\caption{
Meridional cross-sections of $\meanB_r/\Beq$ at different times
for Run~Q2.
}
\label{fig:pa}
\end{figure}

\begin{figure}[t]
\includegraphics[width=0.49\columnwidth]{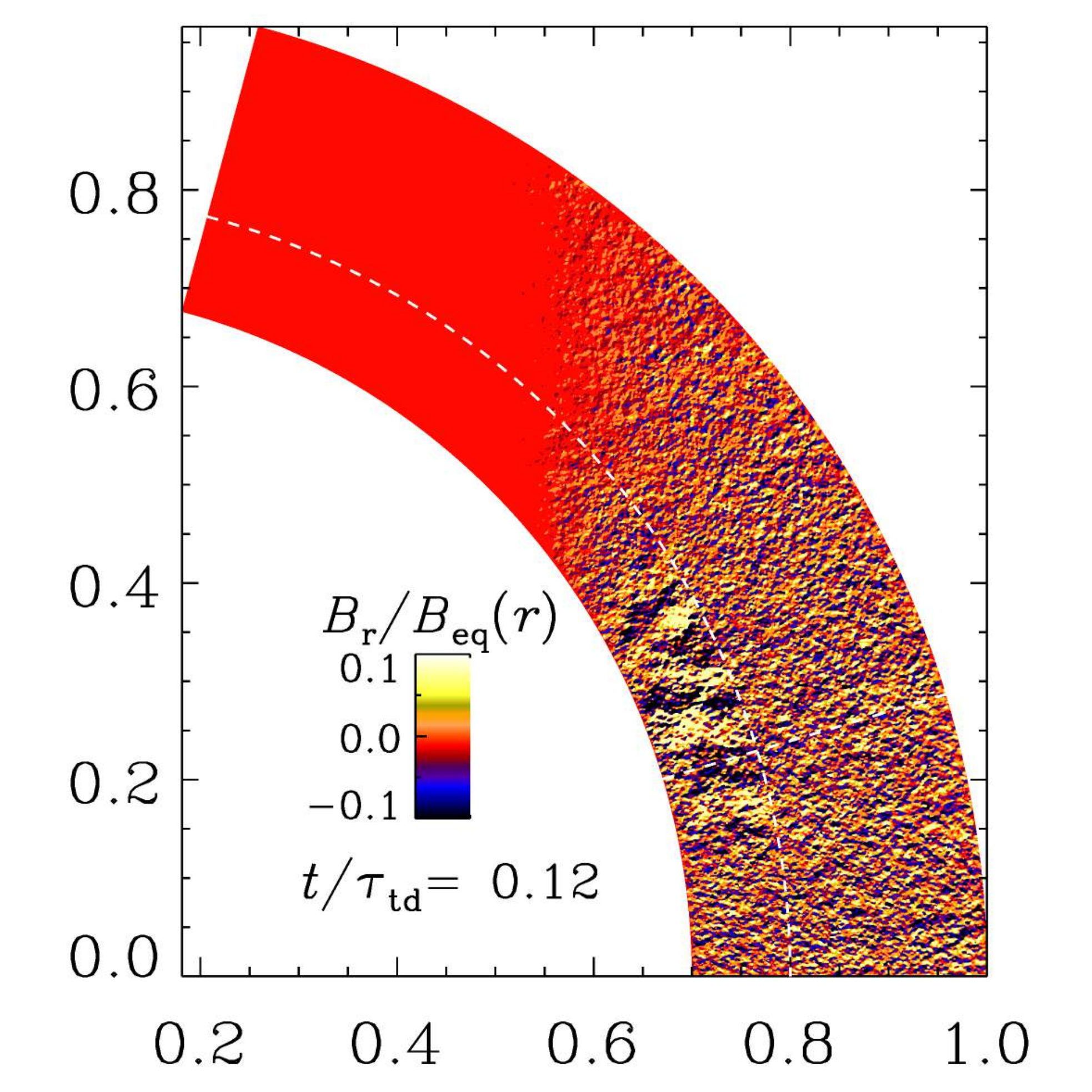}
\includegraphics[width=0.49\columnwidth]{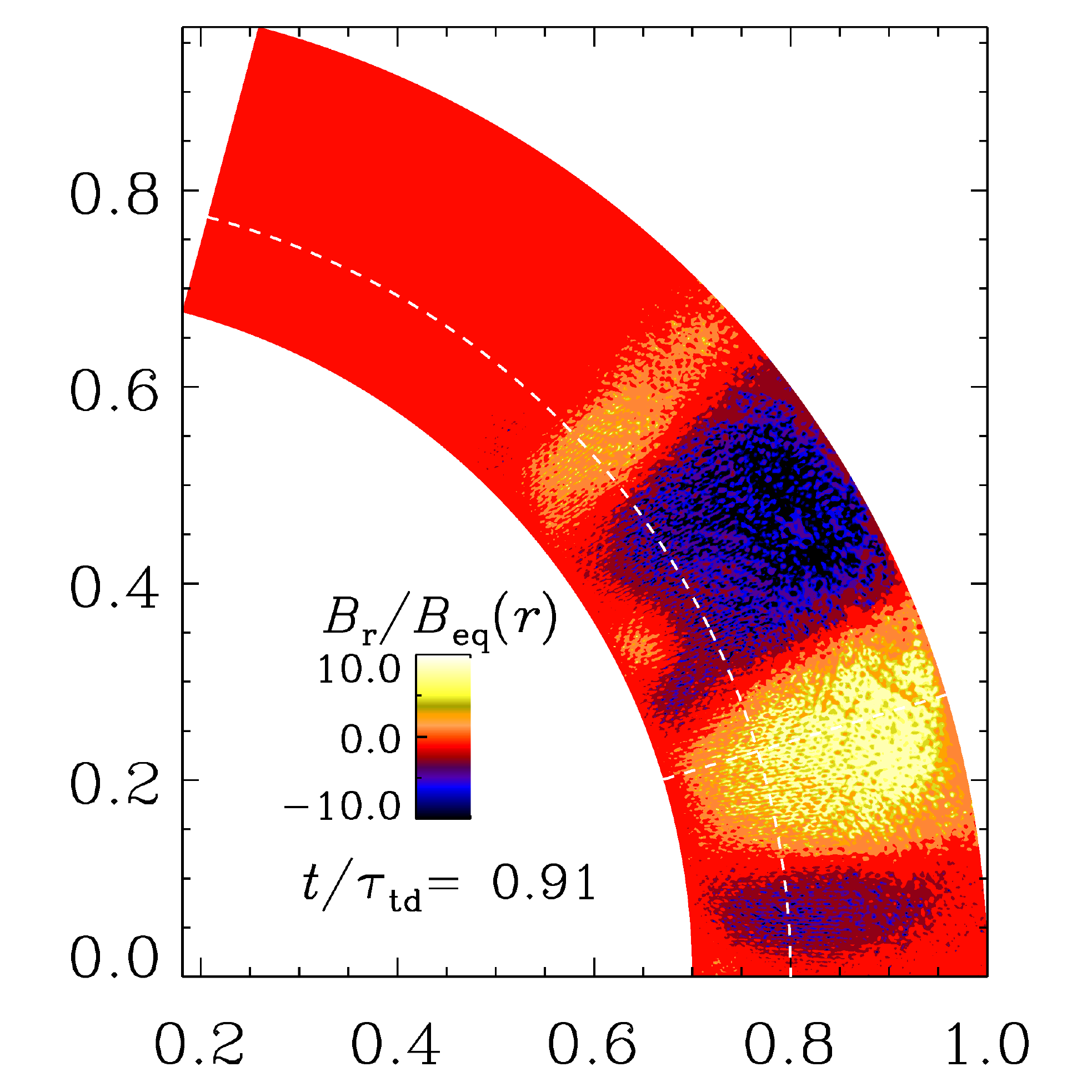}
\caption{
\blue{
Similar to the first and last panels of \Fig{fig:pa},
but the range of the color scale is adapted to the actual extrema.
}
}\label{fig:pb}
\end{figure}

\begin{table}[htb]\caption{
Summary of the runs.
The reference run is shown in bold.
}\vspace{12pt}\centerline{\begin{tabular}{lrlcccccl}
Run & $\Gamma_\rho$ & $\rstar$ & $\phi$ ext. & $\Rm$ & $n$ & b.c.& $\smax$ & $\;\;\tilde\lambda$ \\
\hline
Q1       & 2   & 0.8 &$\pi$  & 20 & 6 &Q &1&0.079\\ %$256dyn_sph_sin_g1_vbc_pi
\blue{Q1b}      & \blue{30}   & \blue{0.8} &\blue{$\pi$}  &\blue{ 20} &\blue{ 6 }&\blue{Q} &\blue{1}&\blue{ 0.085}\\ %$256dyn_sph_sin_g8_vbc_pi
\blue{Q1c}       & \blue{70}   &\blue{ 0.8} &\blue{$\pi$}  &\blue{ 20 }&\blue{ 6} &\blue{Q }&1&\blue{0.083}\\ %$256dyn_sph_sin_g10_vbc_pi
{\bf Q2}&{\bf450}&{\bf0.8}&$\ppi$ &{\bf20}&{\bf6}&{\bf Q}&{\bf1}&{\bf0.084}\\ %$256dyn_sph_sin_g14_vbc_pi
Q3       & 1400& 0.8 &$\pi$  & 20 & 6 &Q &1&\blue{0.074}\\%$256dyn_sph_sin_g17_vbc_pi
Q4       & 450 & 0.75&$\pi$  & 20 & 6 &Q &1&0.074\\%$256dyn_sph_sin_g14_vbc_pi_r75
D1       & 2 & 0.8 &$\pi$  & 20 & 6 &D&1&0.087\\%$256dyn_sph_sin_g1_dip_pi_stress
{\bf D2}       &{\bf450} &{\bf0.8} &$\ppi$  & {\bf20}& {\bf6}& {\bf D}&{\bf1}&{\bf0.082}\\%$256dyn_sph_sin_g14_dip_pi_stress
D3       & 1400 & 0.8 &$\pi$  & 20 & 6 &D&1&0.077\\%$256dyn_sph_sin_g17_dip_pi_stress
D4       & 450 & 0.8 &$\pi$  & 20 & 6 &D&0.2&0.11\\%$256dyn_sph_sin_g14_dip_pi_sig
\blue{D5}       & \blue{450} & \blue{0.8} &$\blue{\pi}$  &\blue{20} & \blue{6} &\blue{D}&\blue{0.5}&\blue{0.11}\\%$256dyn_sph_sin_g14_dip_pi_2x48x64proc_sig05
R1       & 450 & 0.8 &$\pi/4$ & 20 & 6 &D&1&0.0083\\%$256dyn_sph_sin_g14_dip_4phi
R2       & 450 & 0.8 &$\pi/4$ & 40 & 6 &D&1&0.075\\%$256dyn_sph_sin_g14b_dip_4phi
H1       & 2 & 0.8 &$\pi$& 20 & 0 &Q &1&0.087\\ %256dyn_sph_cos_g1_vbc_pi
H2       & 450 & 0.8 &$\pi$& 20 & 0 &Q &1&0.083\\ %256dyn_sph_cos_g14_vbc_pi
H3       & 1400 & 0.8 &$\pi$& 20 & 0 &Q &1&0.076\\ %256dyn_sph_cos_g17_vbc_pi
H4       & 450 & 0.75 &$\pi$& 20 & 0 &Q &1&0.081 %256dyn_sph_cos_g14_vbc_pi_r75
\label{Tab1}\end{tabular}}
\tablecomments{The column `b.c.' indicates whether the equatorial boundary condition is
dipolar (D) or quadrupolar (Q).}
\end{table}

\section{Results}

\subsection{Nature of the dynamo}

Mean-field simulations of $\alpha^2$ dynamos in spherical wedges
have shown that the magnetic field is strongest near the high-latitude
boundaries \citep{Jab13}.
However, in rapidly rotating stratified spherical shell convection,
the kinetic helicity is typically found to be maximum close to the equator,
e.g.\ at $\pm15\degr$ latitude \citep{KMB12}.
For this reason, we focus in the present paper on the case $n=6$,
which yields a maximum of the magnetic field at about $22\degr$.

In \Fig{fig:butt} we show butterfly diagrams of $B_r(\theta, t)/\Beq$
at $r/R=0.75$, $0.8$, and $0.95$ and $\phi=0$.
We observe equatorward migration of the magnetic field at low latitudes
(below $20\degr$) and poleward migration at higher ones.
\blue{
Note, however, that the latitudinal variation of the magnetic field
is much more complex than the field variations of similar mean-field
calculations \citep{Jab13} and even DNS with forced unstratified
turbulence \citep{WBM11}.
A possible reason for this can be the larger aspect ratio of the
dynamo-active layer, which is now rather thin.
This can lead to a larger number of toroidal flux belts \citep{MTB90}.
Another reason could be the comparatively short run time ($6.5\tautd$),
which might imply that the field is still in a transient.
However, in view of the comparatively large spatial resolution
($256\times1152\times1152$
mesh points), longer runs become computationally prohibitive.
Note also the occurrence of sharp structures in the bottom panel of
\Fig{fig:butt} at $t/\tautd\approx5.5$ (and also $\approx3.5$).
This rapid time variation is a consequence of plotting the field
at a fixed values of $\phi$ (here $\phi=0$)
and the fact that the non-axisymmetric structures drift in $\phi$
(here in the westward direction).
}

\begin{figure*}[t]
\centering
\hspace*{-1mm}\includegraphics[width=0.25\textwidth]{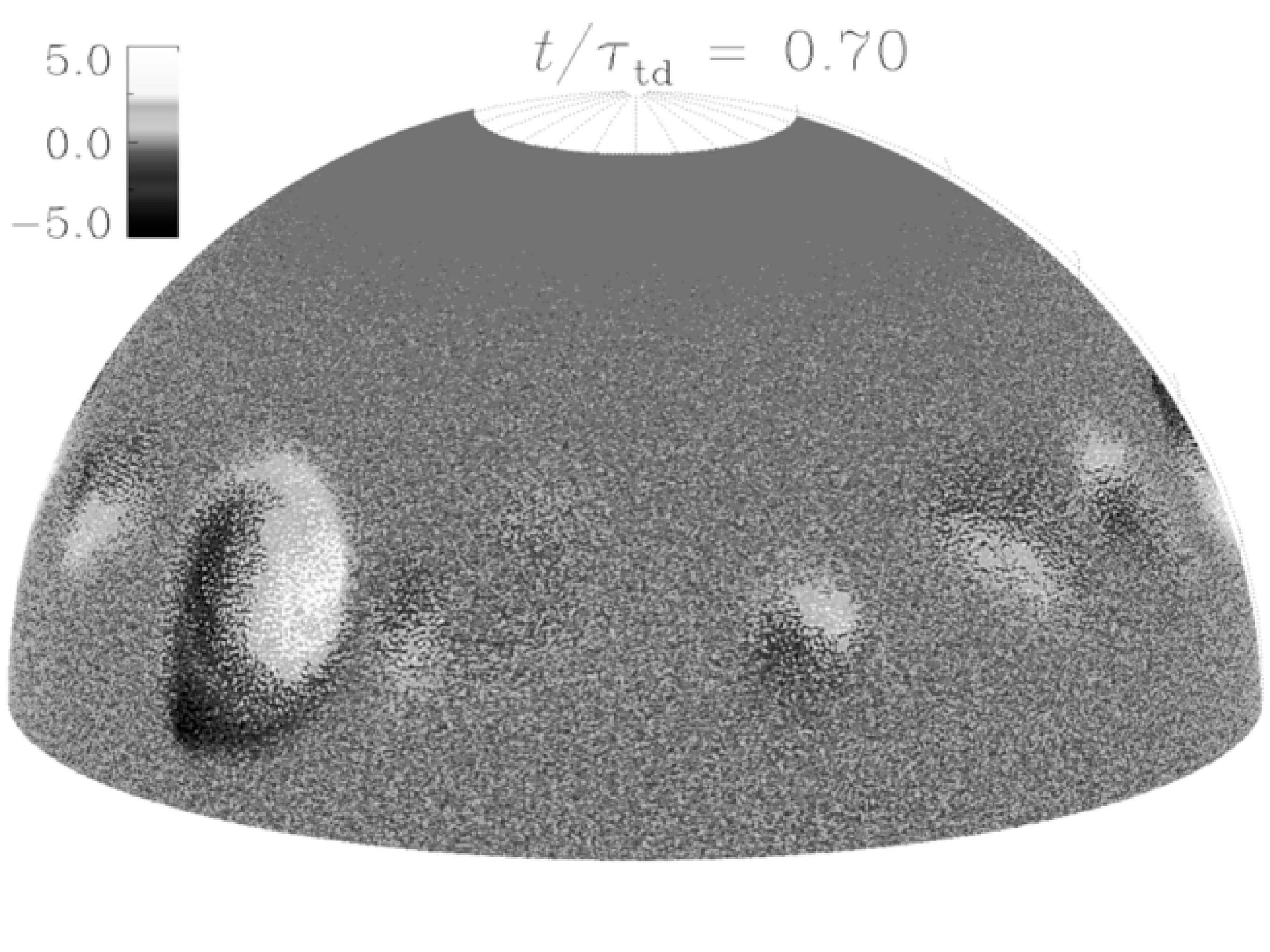}
\hspace*{-1mm}\includegraphics[width=0.25\textwidth]{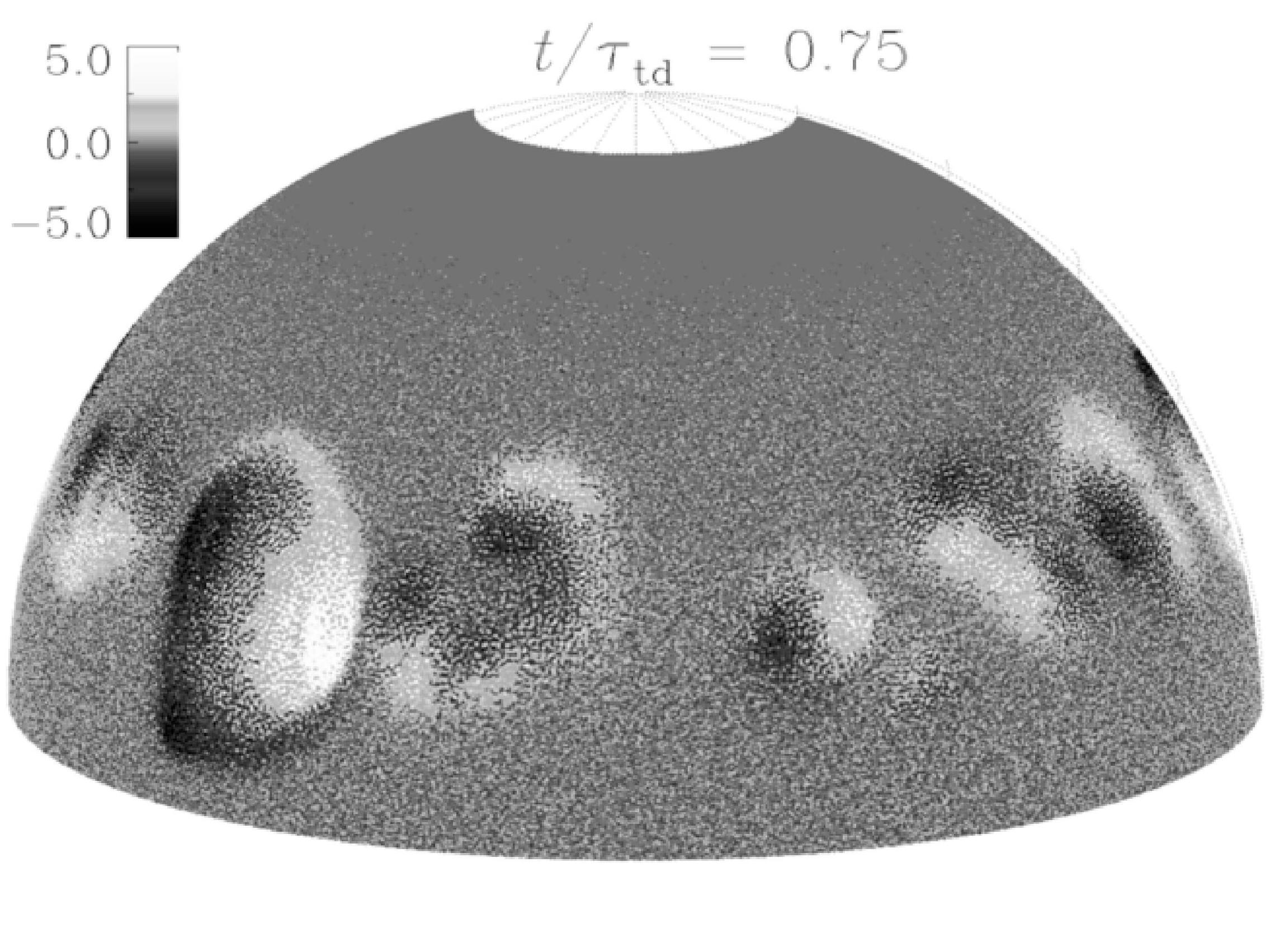}
\hspace*{-1mm}\includegraphics[width=0.25\textwidth]{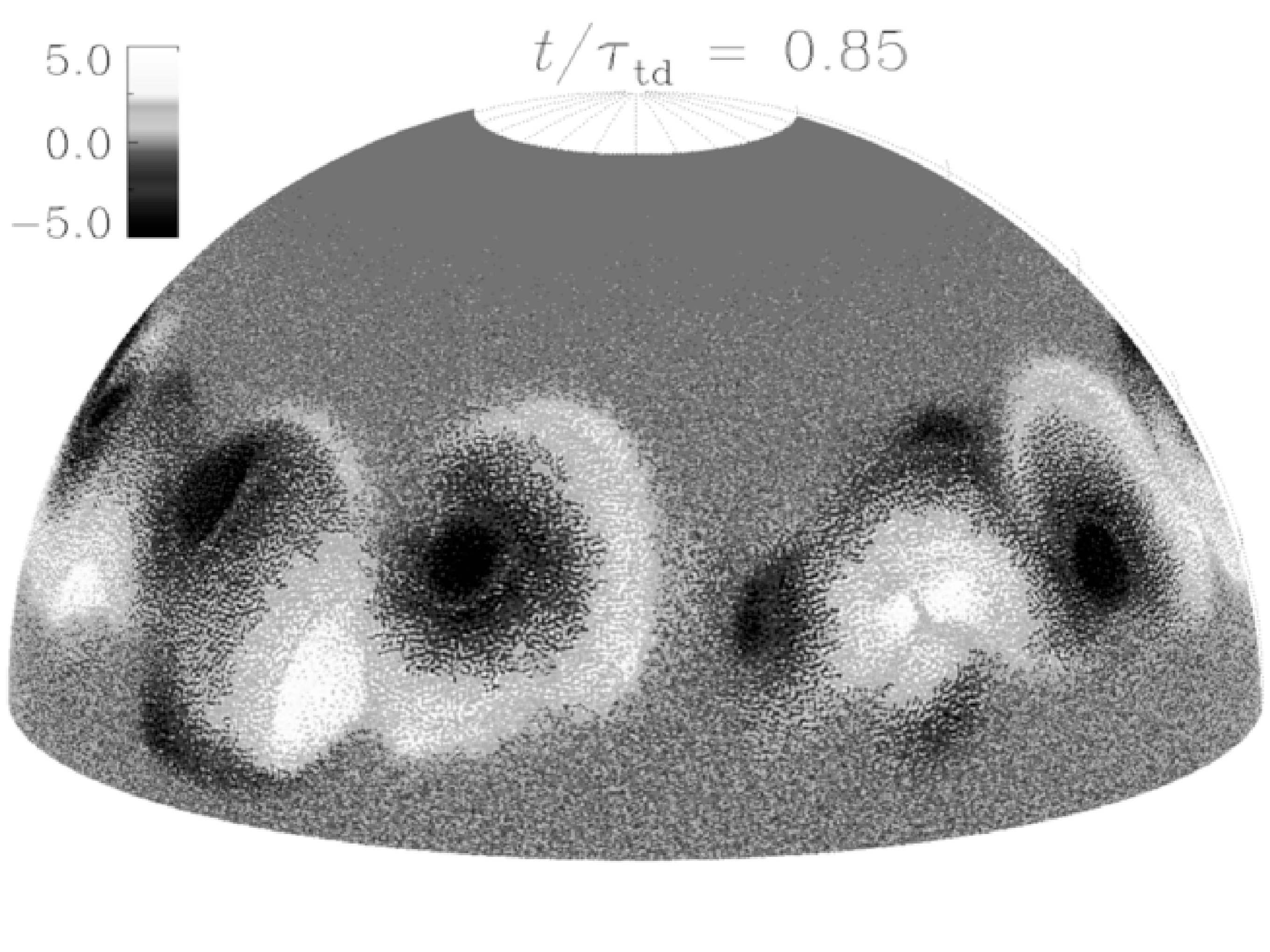}
\hspace*{-1mm}\includegraphics[width=0.25\textwidth]{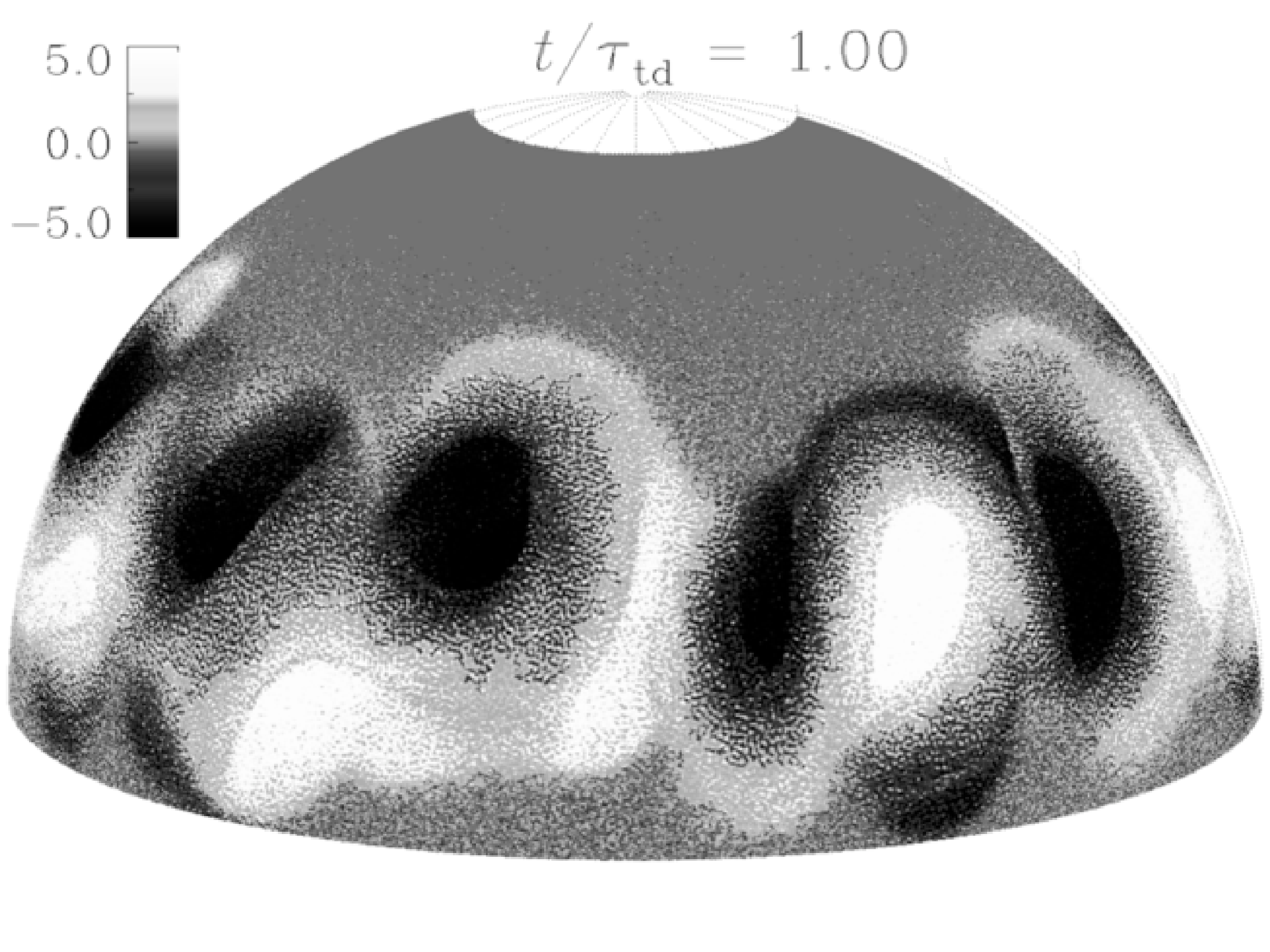}\\
\caption{
Time evolution of $B_r/\Beq$ at $r/R=0.98$ for a simulation
with $\Gamma_\rho=450$ for Run~Q2.
%\url{256dyn_sph_sin_g14_vbc_pi}
}\label{fig:pptQ}
\end{figure*}

\begin{figure*}[t]
\centering
\hspace*{-1mm}\includegraphics[width=0.25\textwidth]{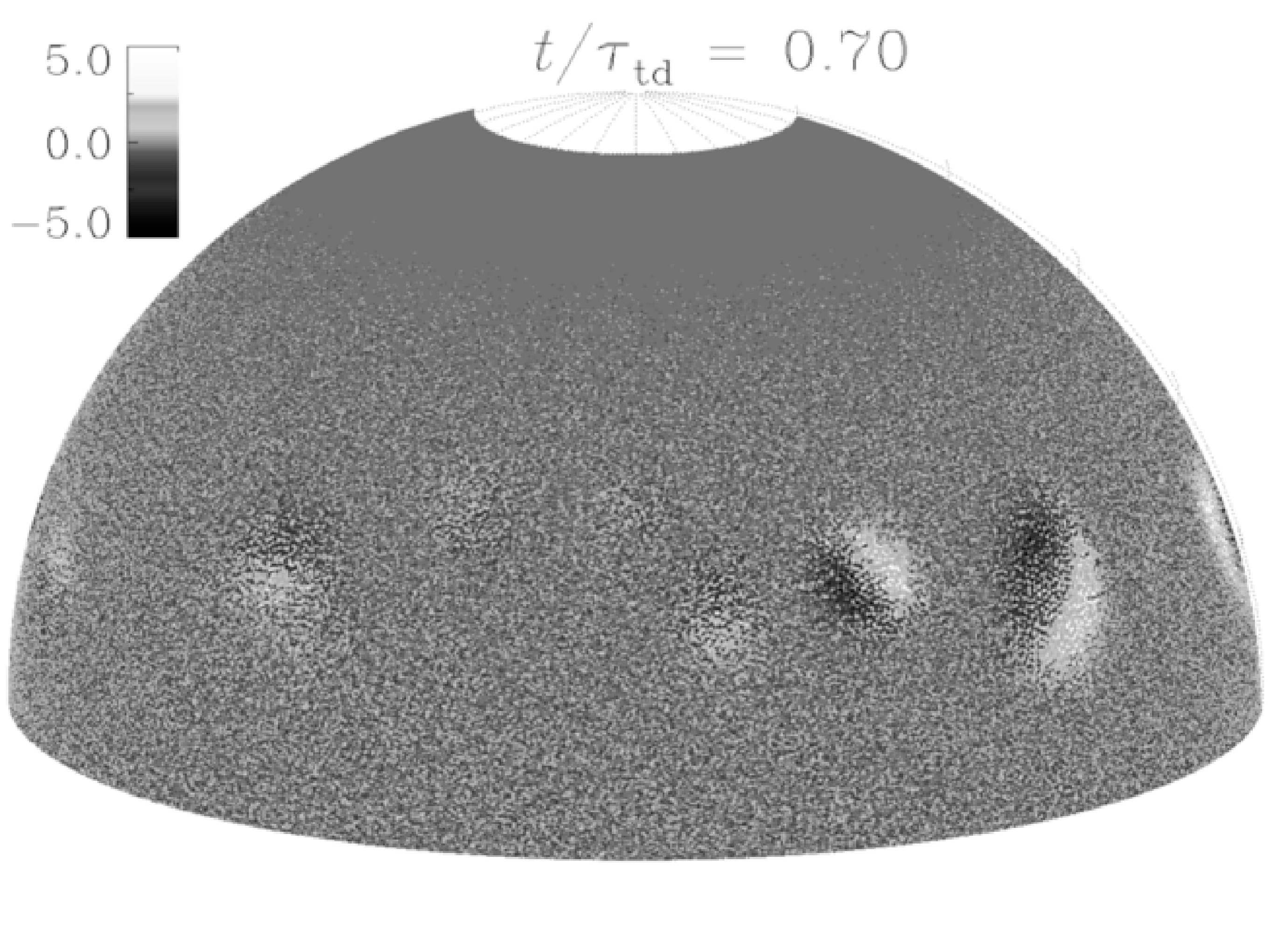}
\hspace*{-1mm}\includegraphics[width=0.25\textwidth]{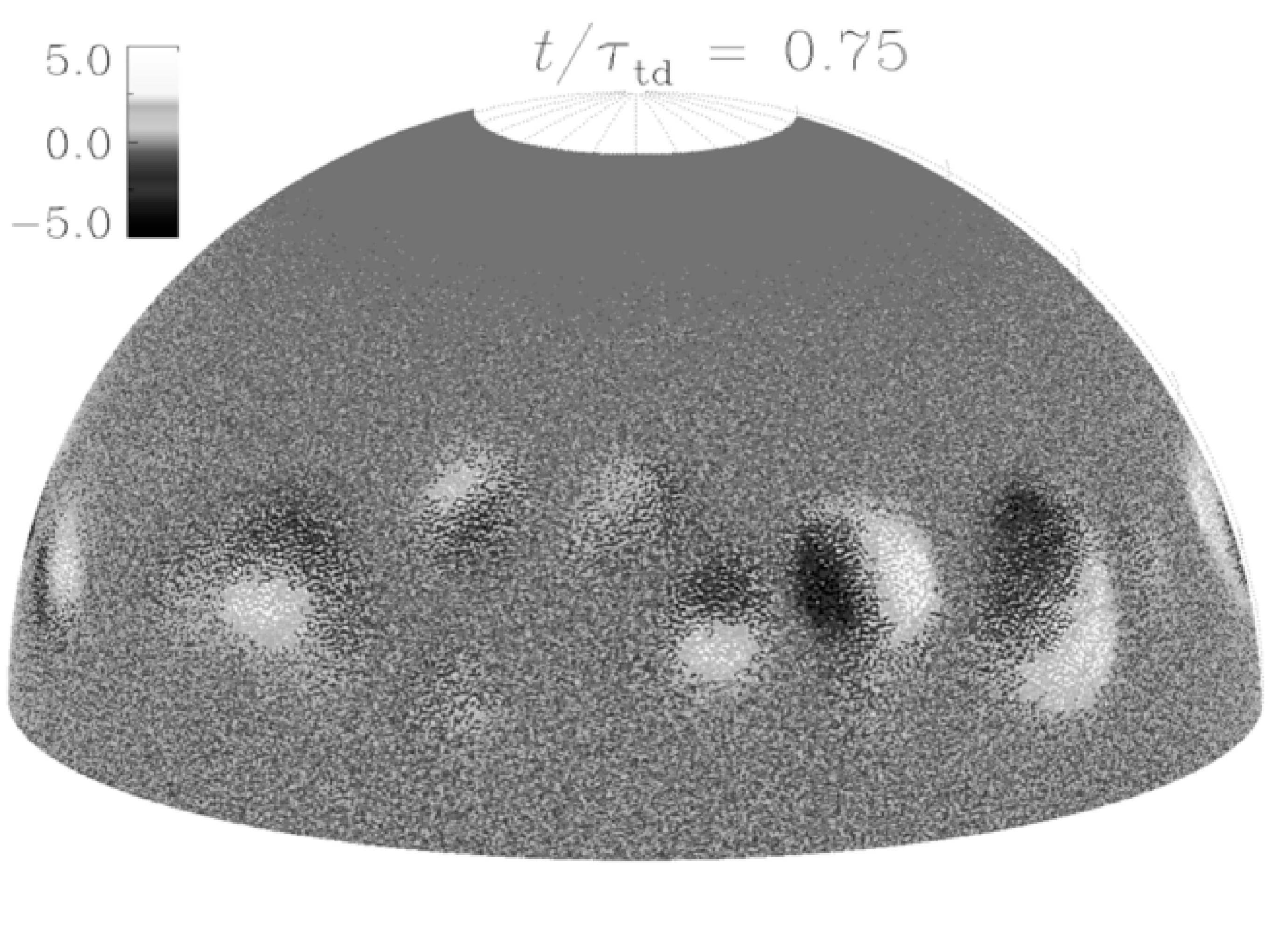}
\hspace*{-1mm}\includegraphics[width=0.25\textwidth]{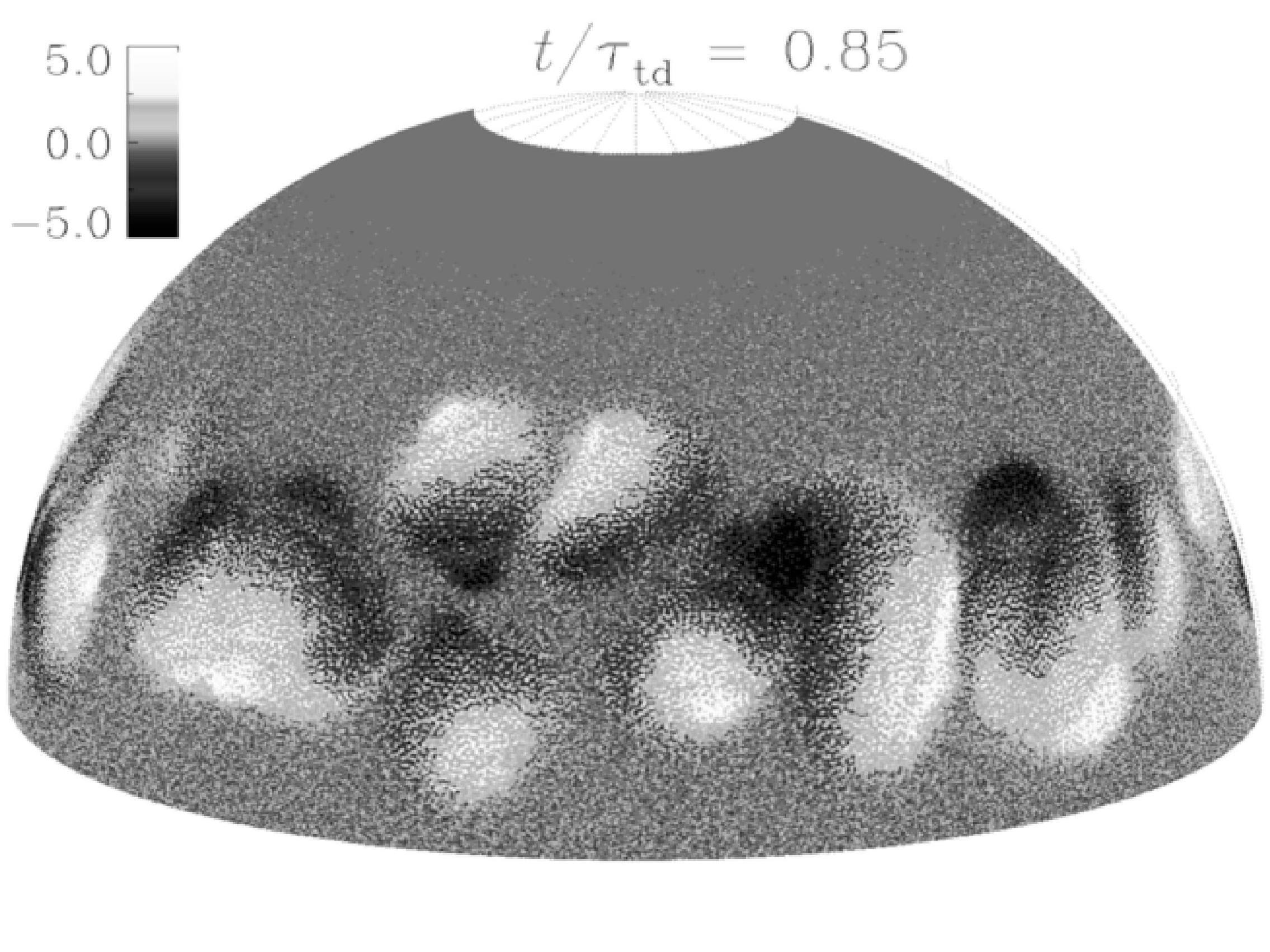}
\hspace*{-1mm}\includegraphics[width=0.25\textwidth]{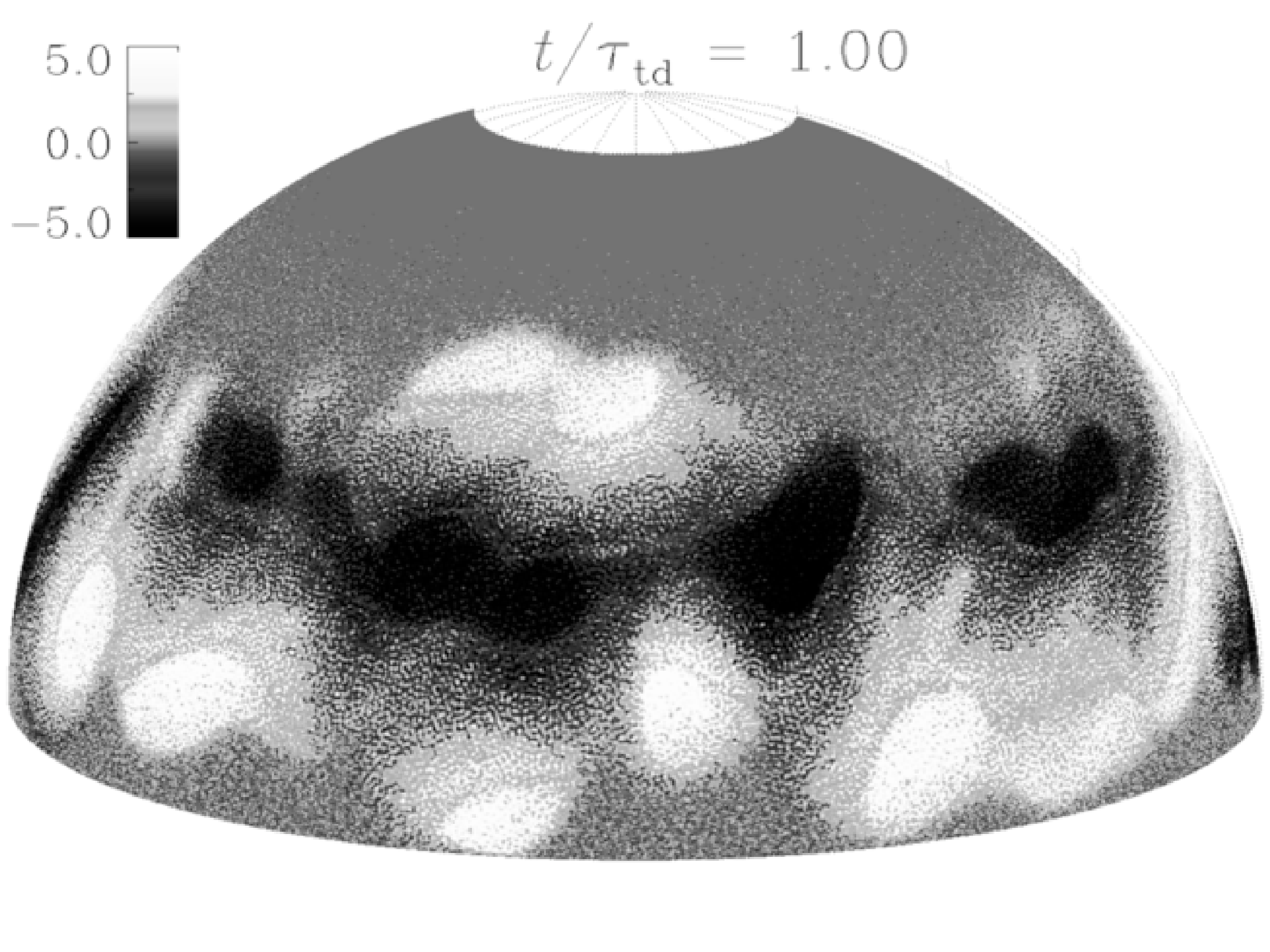}\\
\caption{
Same as \Fig{fig:pptQ}, but for Run~D2.
%\url{256dyn_sph_sin_g14_dip_pi}
}\label{fig:pptD}
\end{figure*}

The observed dynamo wave is generated by an $\alpha^2$ dynamo.
The fact that $\alpha^2$ dynamos with nonuniform $\alpha$ distribution can be
oscillatory was known for some time \citep{shu+sok+ruz85,bar+shu87,ste+ger04},
but their migratory properties were first pointed out by \cite{Mit10}.
Not surprisingly, the magnetic field is generated in the lower layer,
as can be seen from \Fig{fig:pa} where we show meridional cross-sections
of $B_r(r,\theta)/\Beq(r)$ at $\phi=0$ at six different times.
\blue{
In this sequence we have chosen a fixed color scale that saturates
at the equipartition level.
This allows us to see at which times and heights the field reaches
equipartition.
However, to see the spatial variation of the field,
we show in \Fig{fig:pb} the first and last times with a color range
that saturates at 0.1 and 10 times the equipartition value.
}

We note that during the early growth phase of the dynamo, the magnetic
field grows exponentially and the non-dimensional growth rate is of the
order of 0.1 (in units of the inverse turbulent diffusion time).
It increases with magnetic Reynolds number (cf.\ Runs~R1 and R2),
which might be related to the possibility of small-scale dynamo action.
This is supported by the fact that the growth rate is not increasing
with helicity (cf.\ Runs~D2 and D4).

\subsection{Spot formation}
\label{SpotFormation}

Next, we consider the surface appearance of the radial magnetic field.
We see the formation of structures at low latitudes
in less than a turbulent diffusive time.
At first a few bipolar regions form.
As time goes on, these structures move, rotate, and expand,
and after a long enough time they form a strong field concentration which
move toward the equator and forms three band-like structures
with opposite polarities (see \Figs{fig:pptQ}{fig:pptD}).
A similar behavior was also observed by \cite{Mit14}; see Figures~3 and 4 of their paper.

\Figs{fig:pptQ}{fig:pptD} illustrate the time evolution of our reference simulations
(Runs~Q2 and D2) with $\Gamma_\rho=450$.
As one can see, at early times of bipolar spot formation, the two polarities are very close
to each other.
One sees that each polarity consists of a core with strong field
and a shadow around it with weaker field.
As time elapses, both core and the shadow expand but
the speed of the expansion of the shadow is larger than
the speed of expansion of the core.
This implies that two polarities start moving apart form each other.
The rest of the evolution is somewhat different for Runs~Q2 and D2.
For Run~Q2, the bipolar spot orientation is preferentially in
the azimuthal direction, both at early and later times.
For Run~D2, on the other hand, the spots tilt in such a way that
the part of the structure with the same polarity tends to occupy a
certain latitudinal band, while that of the opposite orientation
occupies a band at a different latitude; see \Fig{fig:pptD}.
This global structure remains without change, no matter how much time passes.

\begin{figure}[t]
\centering
\hspace*{-1mm}\includegraphics[width=0.5\textwidth]{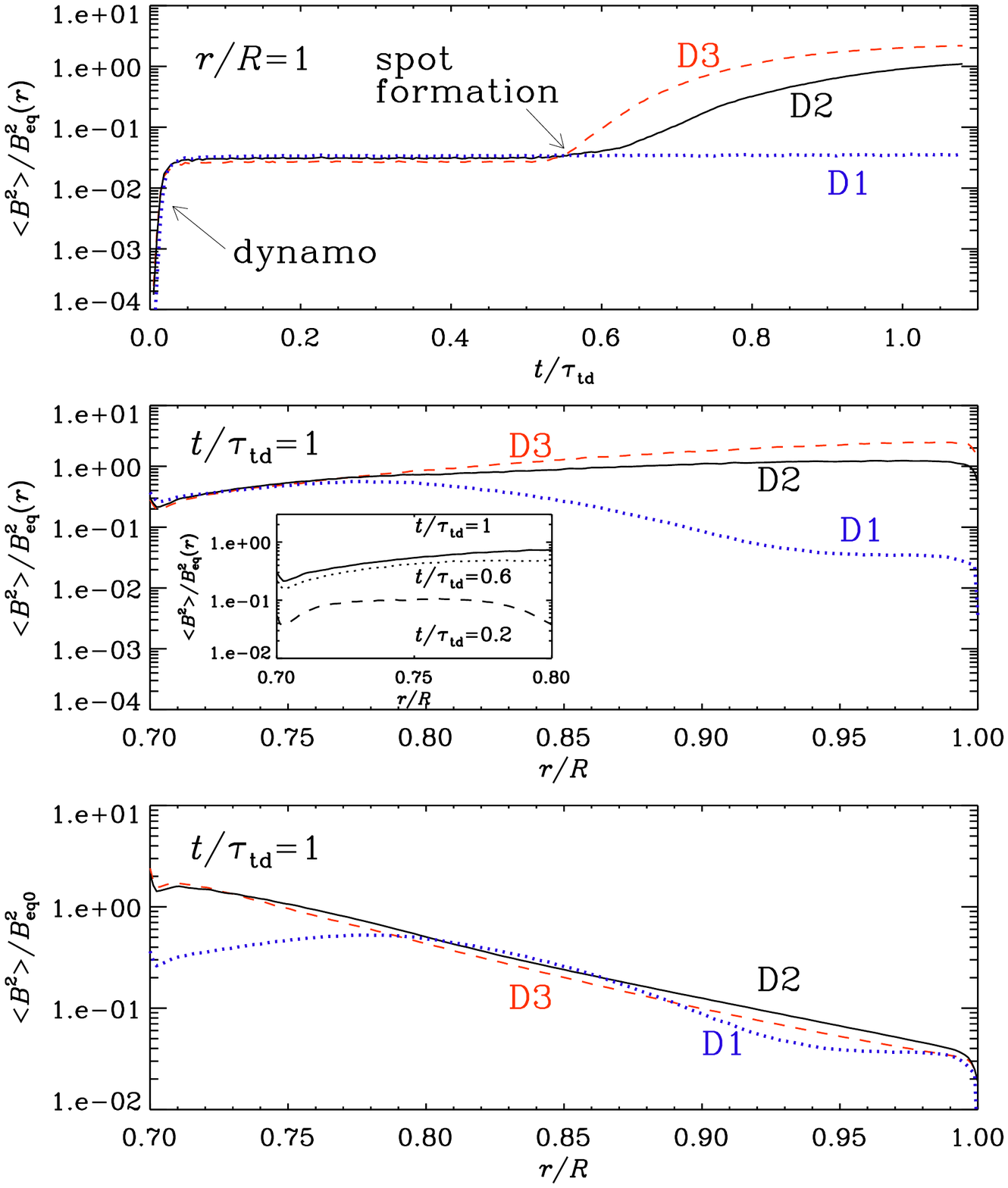}
\caption{
Upper panel: time evolution of $\bra{\BB^2}/\Beq^2(r)$
for Runs~D1 (dotted, blue), D2 (solid, black), and D3 (dashed, red).
Middle panel: radial dependence of $\bra{\BB^2}/\Beq^2(r)$ at $t/\tautd=1$
for Runs~D1--D3.
\blue{
The inset shows the radial dependence of $\bra{\BB^2}/\Beq^2(r)$
for Run~D2 inside the $r/R<0.8$ (helical zone)
at $t/\tautd=0.2$ (dotted), $0.6$ (dashed), and $1$ (solid).
}
\blue{
Lower panel: radial dependence of $\bra{\BB^2}/\Beqz^2$ for
fixed normalization.
}
%\url{256dyn_sph_sin_g1_vbc_pi}
%\url{256dyn_sph_sin_g14_vbc_pi}
%\url{256dyn_sph_sin_g17_vbc_pi}
}
\label{fig:keme}
\end{figure}

\begin{figure}[t]
\centering
\includegraphics[width=\columnwidth]{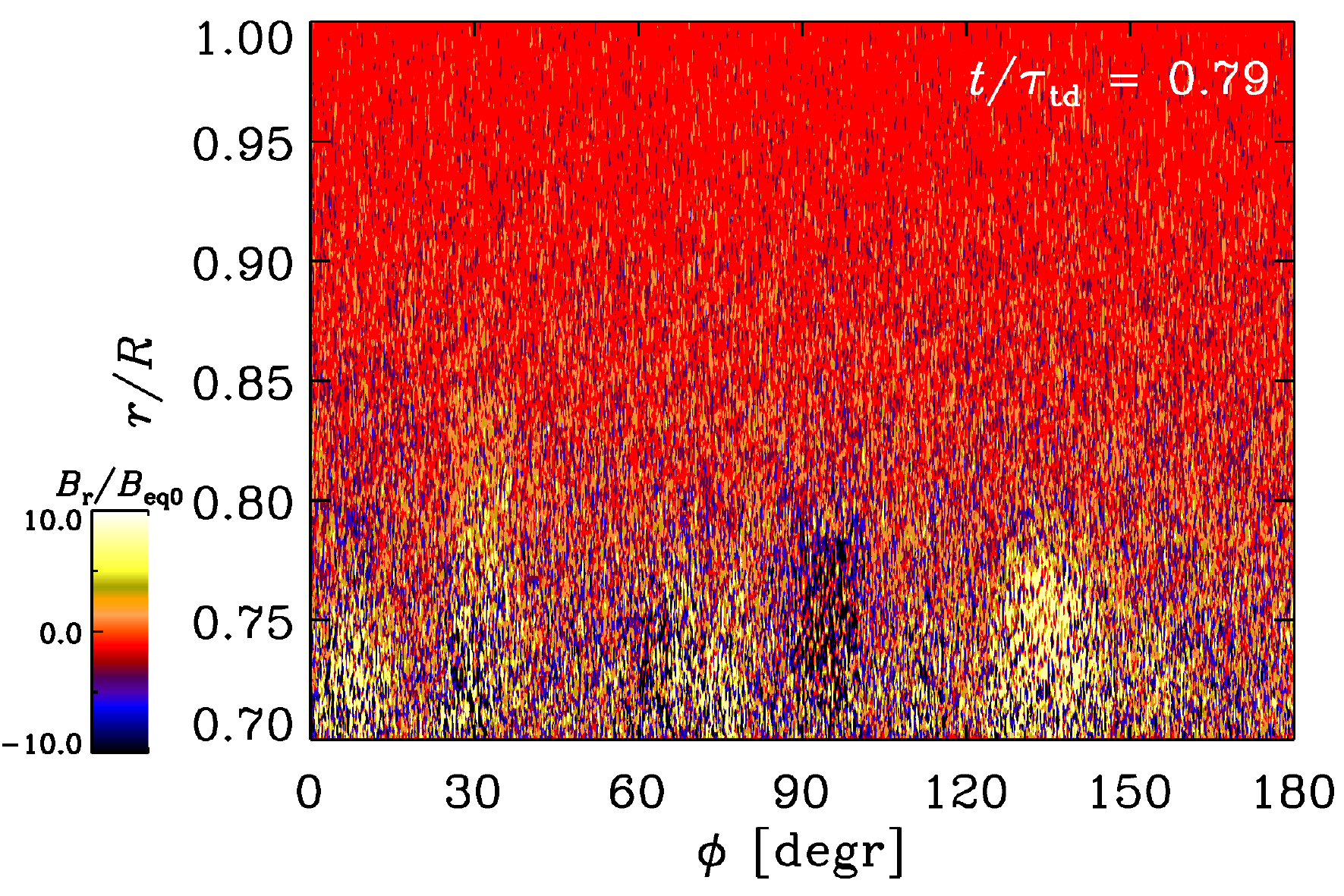}
\includegraphics[width=\columnwidth]{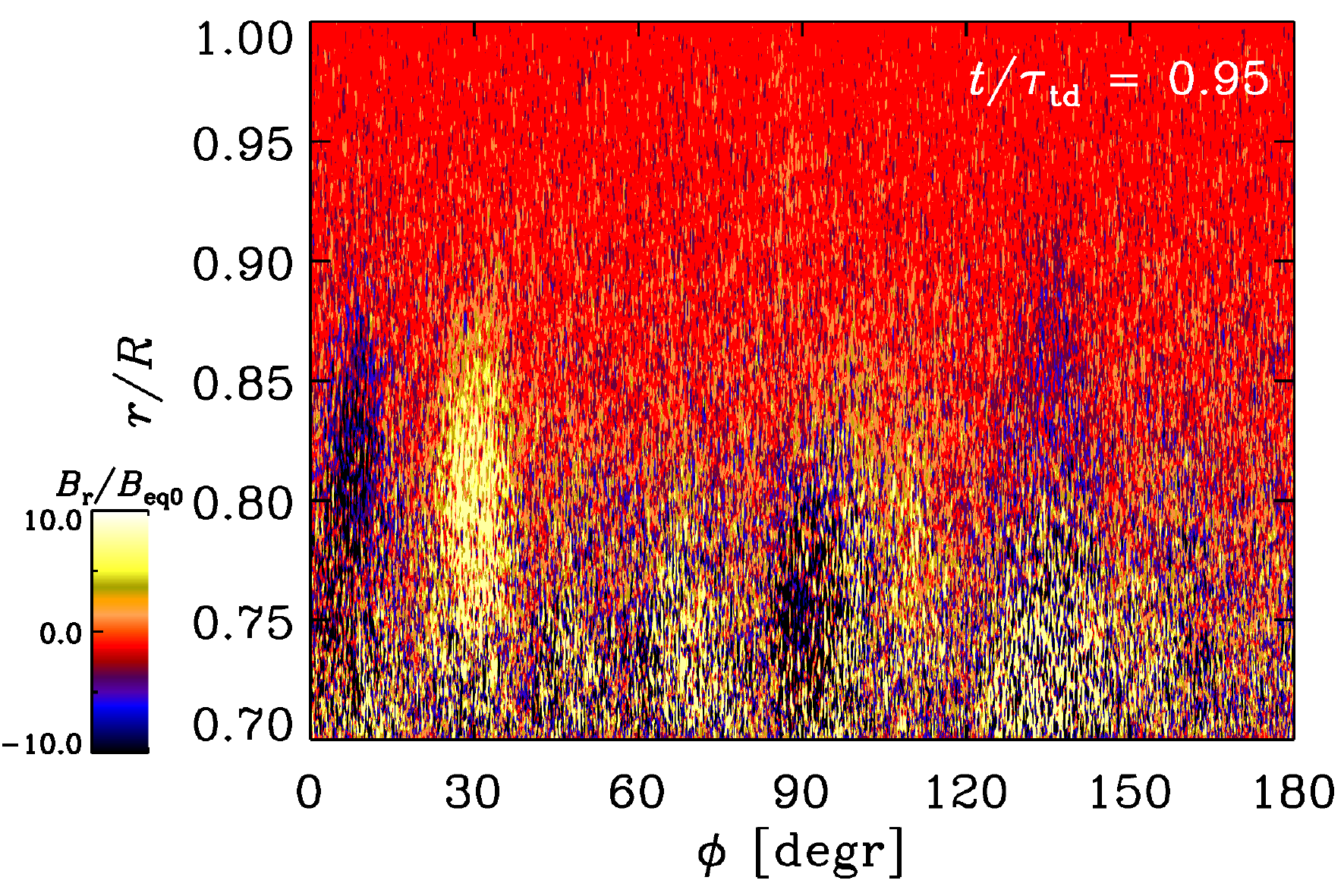}
\includegraphics[width=\columnwidth]{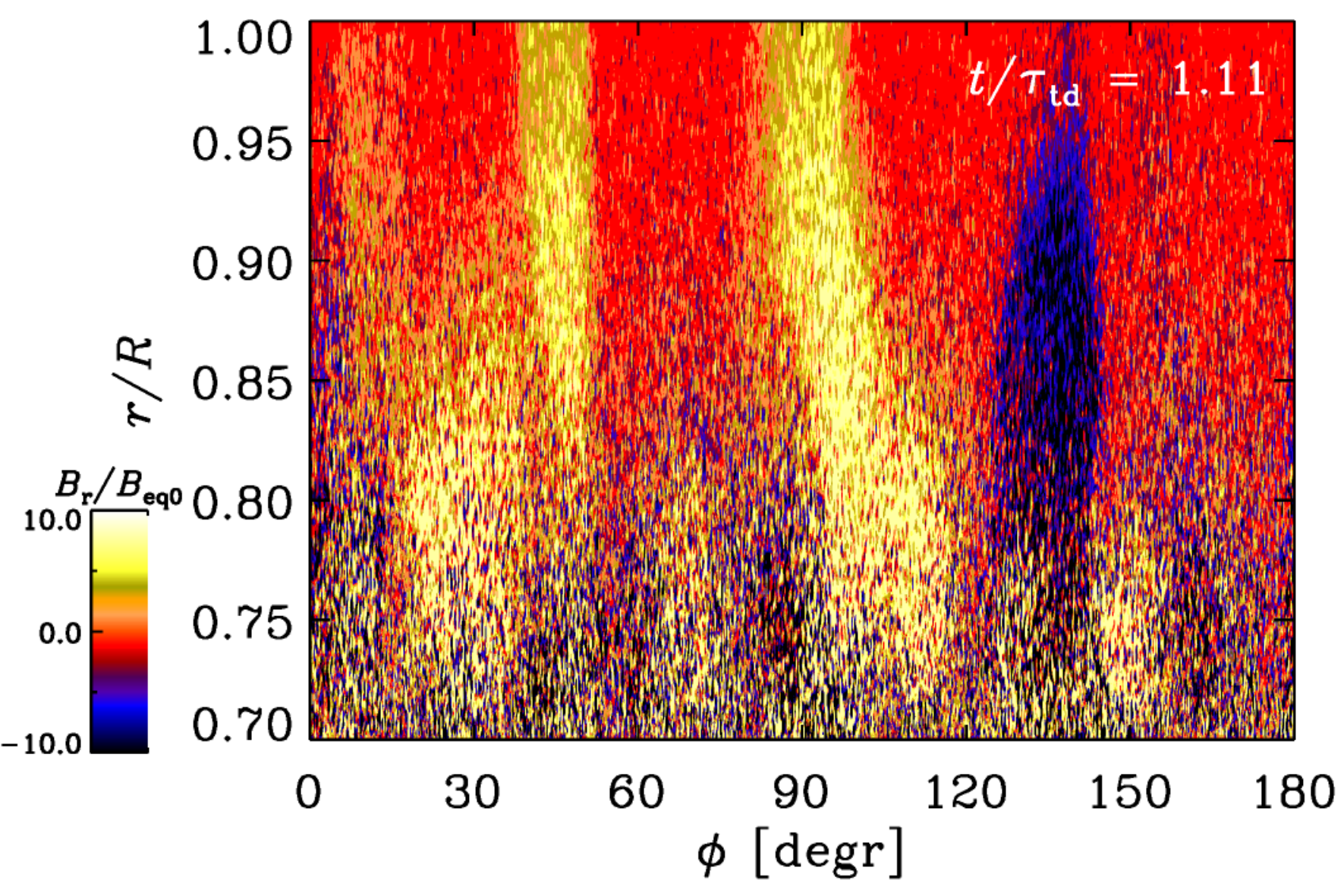}
\caption{
$r\phi$ projection of $B_r/\Beqz$ along the active region belt
for Run~D2.
}\label{fig:xz_slice}
\end{figure}

\begin{figure}[t]
\centering
\includegraphics[width=\columnwidth]{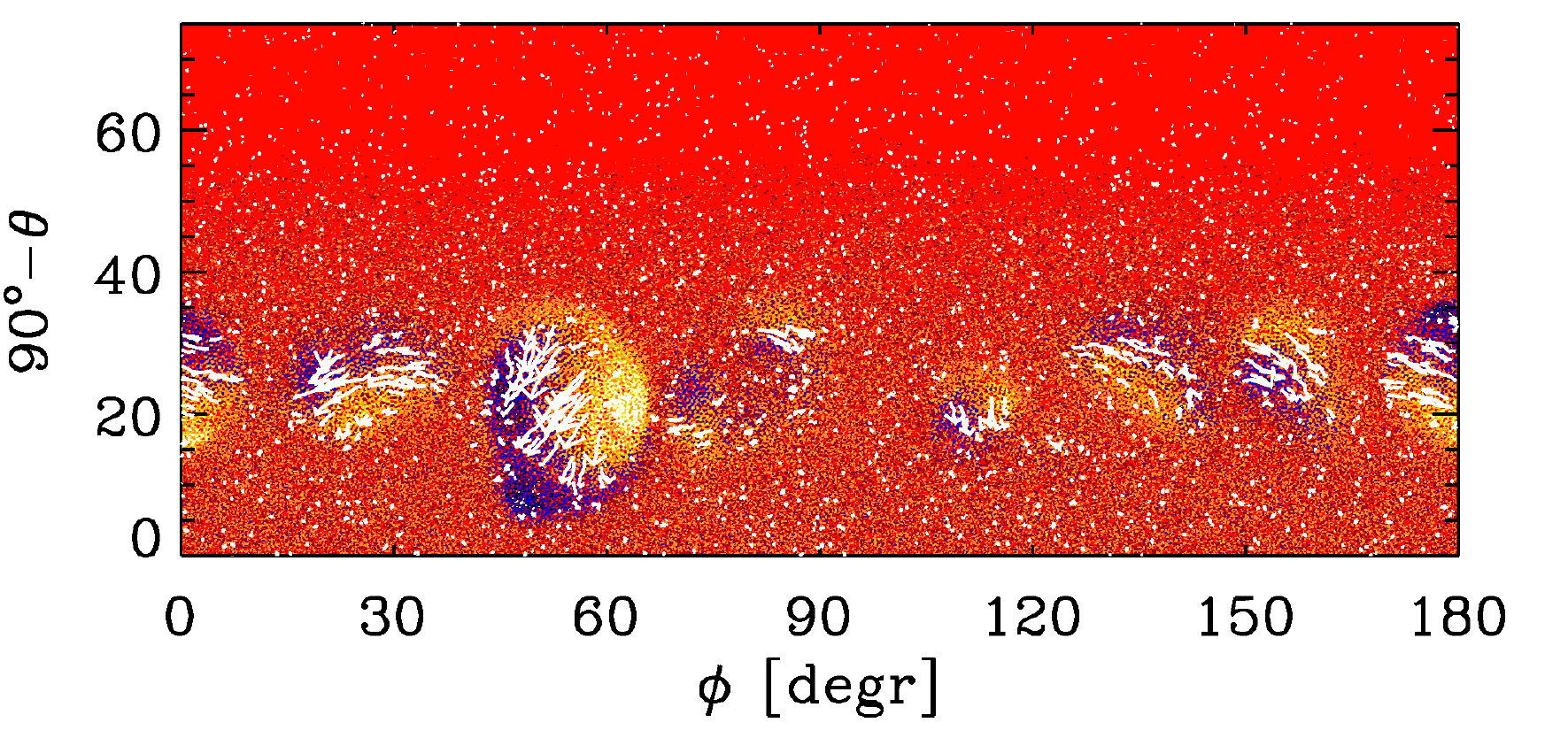}
\caption{
Mercator projection of $B_r/\Beq$ with $(B_\theta,B_\phi)$ vectors superimposed
just below the surface at $t/\tautd=0.7$ for Run~Q2.
}\label{fig:yz_slice30}
\end{figure}

\begin{figure*}[t]
\centering
\hspace*{-1mm}\includegraphics[width=0.3\textwidth]{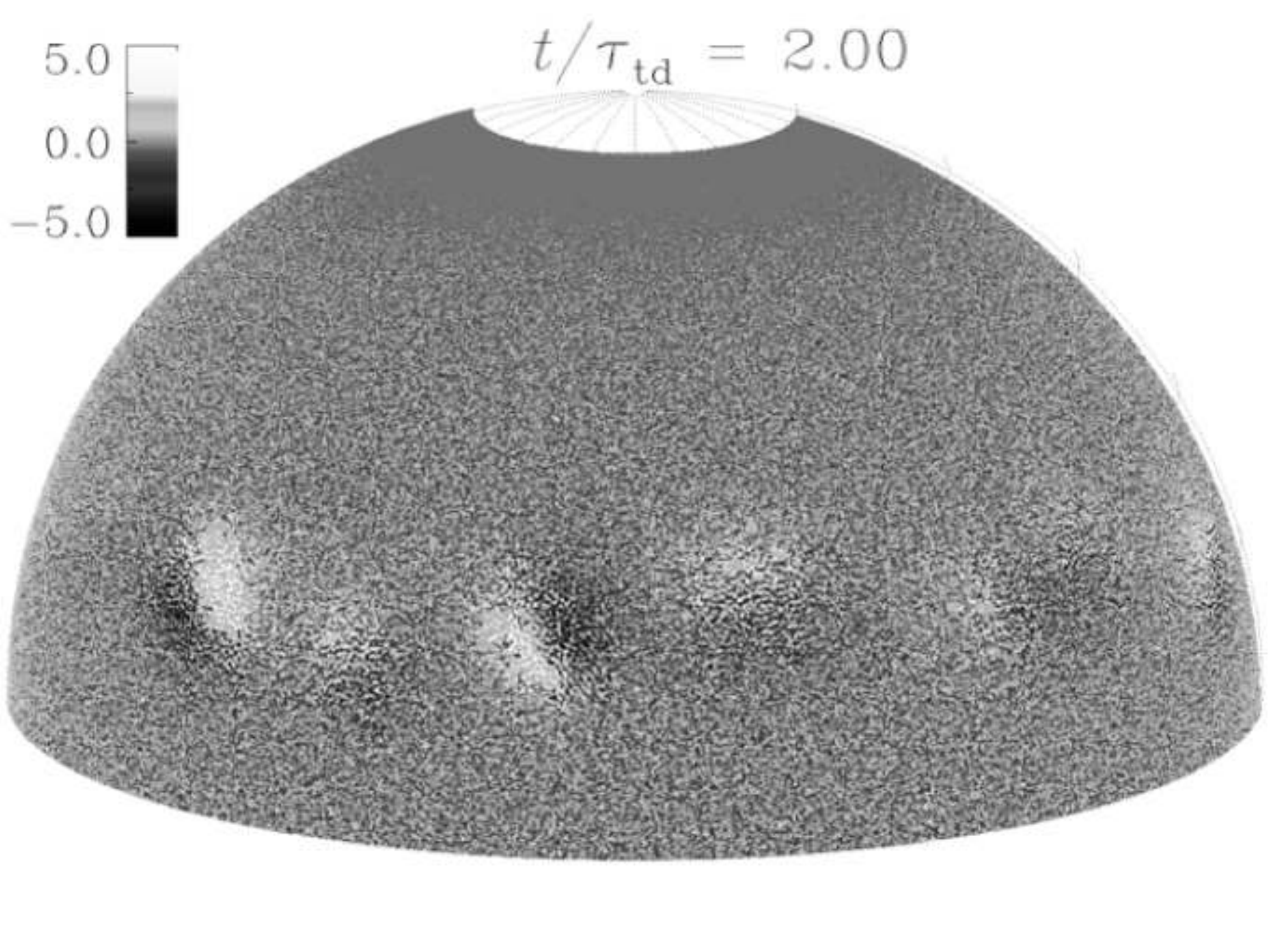}
\hspace*{-1mm}\includegraphics[width=0.3\textwidth]{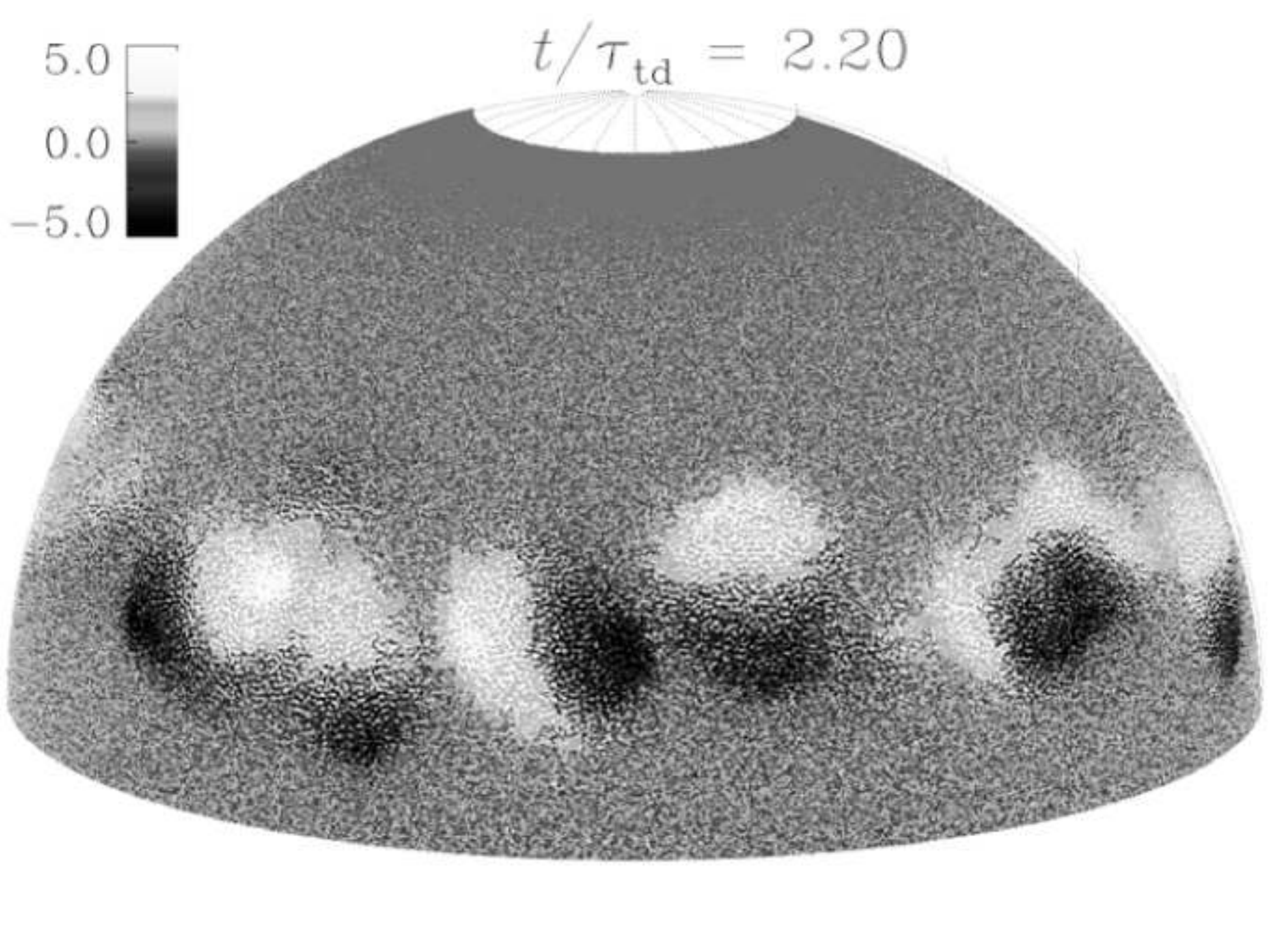}
\hspace*{-1mm}\includegraphics[width=0.3\textwidth]{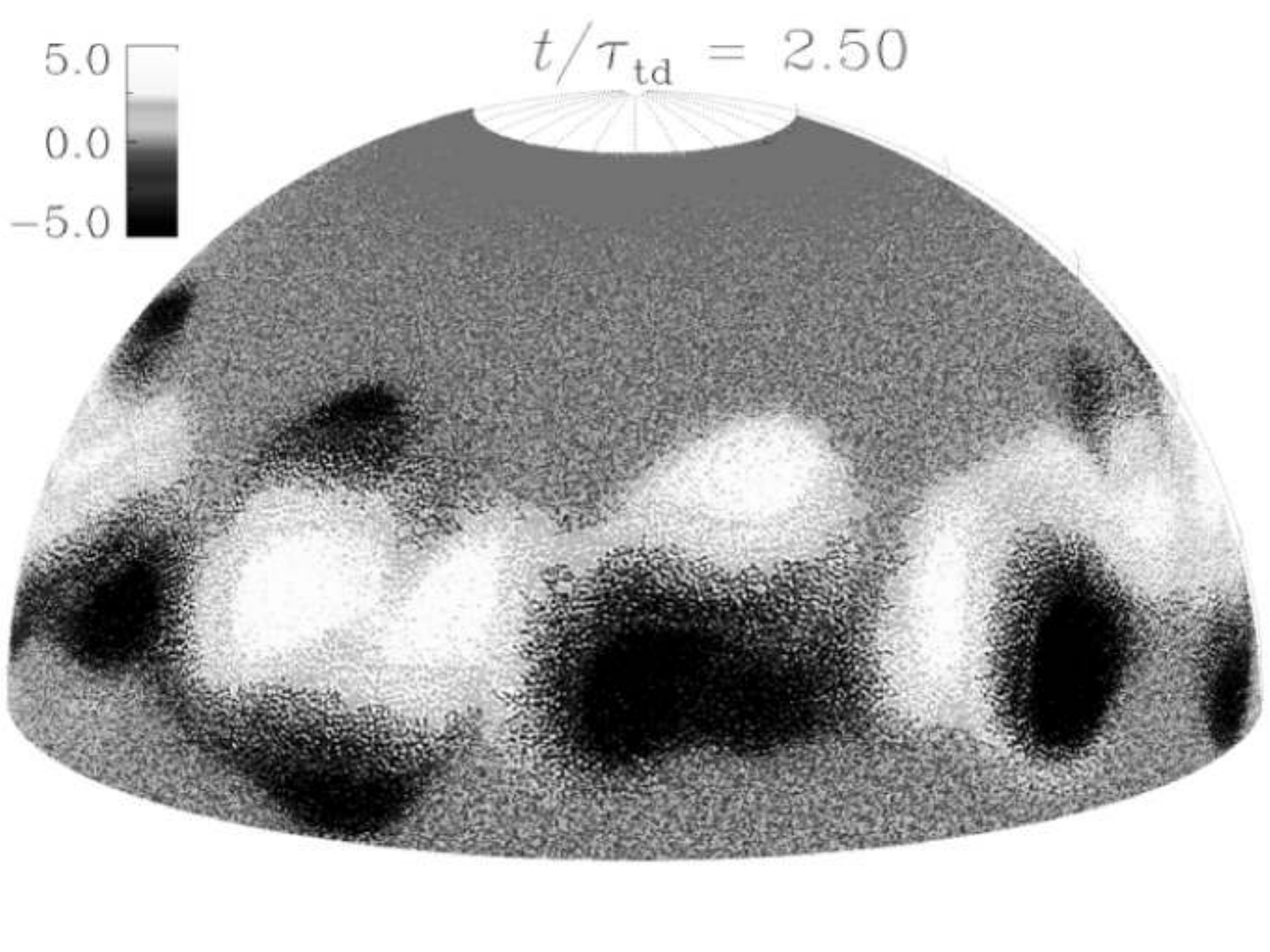}
\caption{
\blue{
Time evolution of $B_r/\Beq$ at $r/R=0.98$ for a simulation
%with $\sigma=0.5$ (Run~D5).
%AB: should be smax
with $\smax=0.5$ (Run~D5).
}
%\url{256dyn_sph_sin_g14_vbc_pi}
}\label{fig:psig}
\end{figure*}

\begin{figure*}[t]
\centering
\hspace*{-1mm}\includegraphics[width=0.3\textwidth]{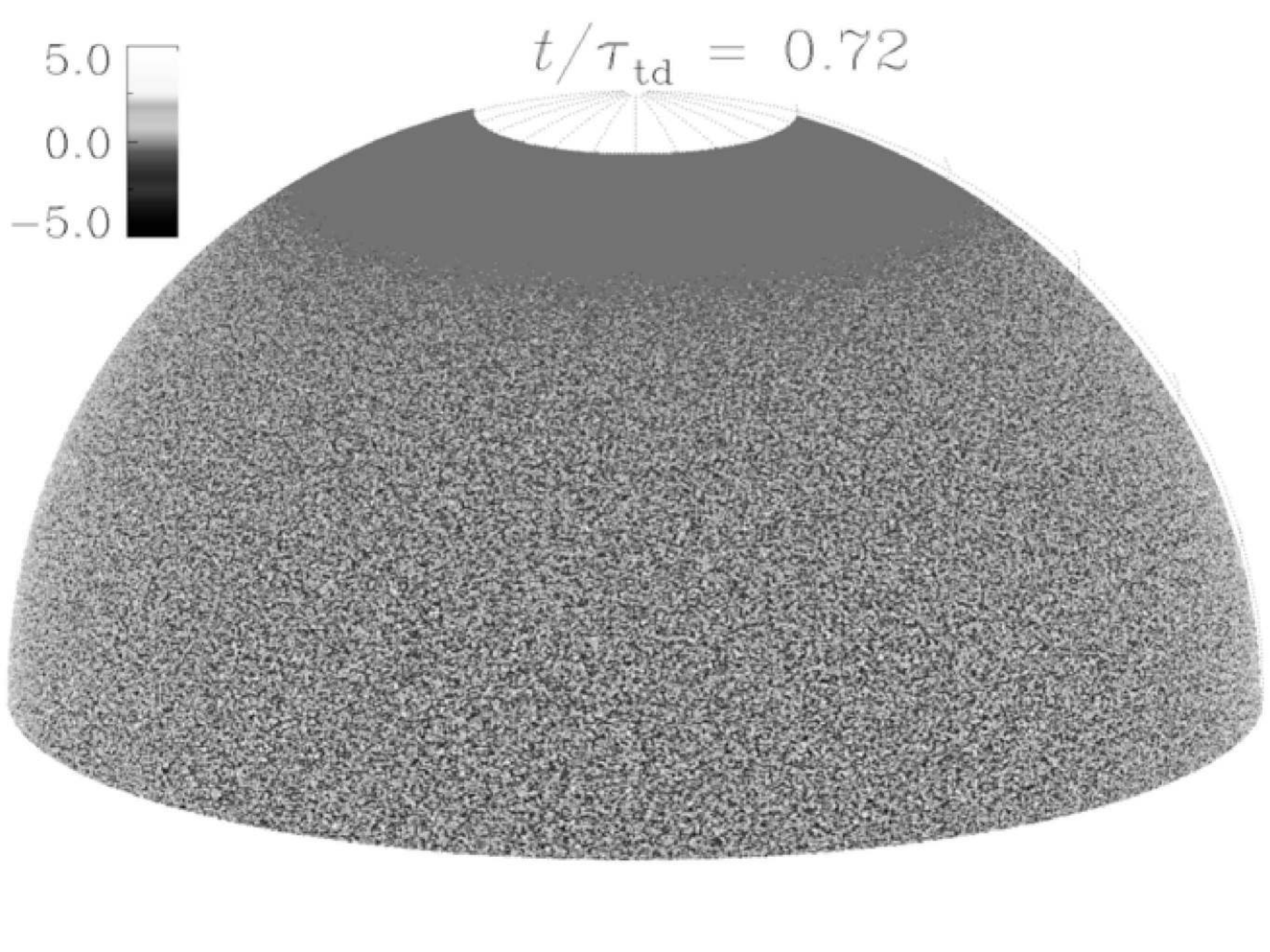}
\hspace*{-1mm}\includegraphics[width=0.3\textwidth]{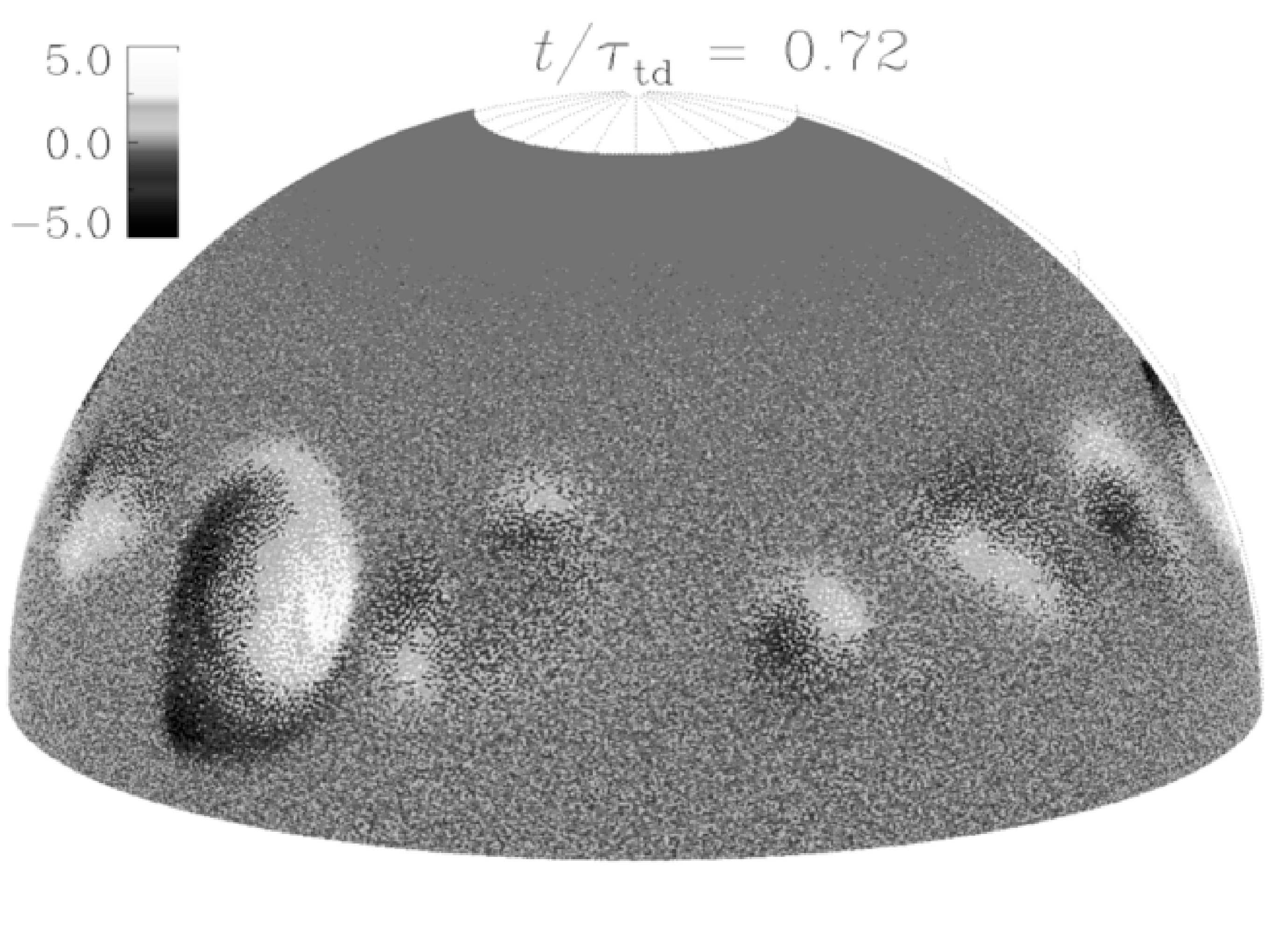}
\hspace*{-1mm}\includegraphics[width=0.3\textwidth]{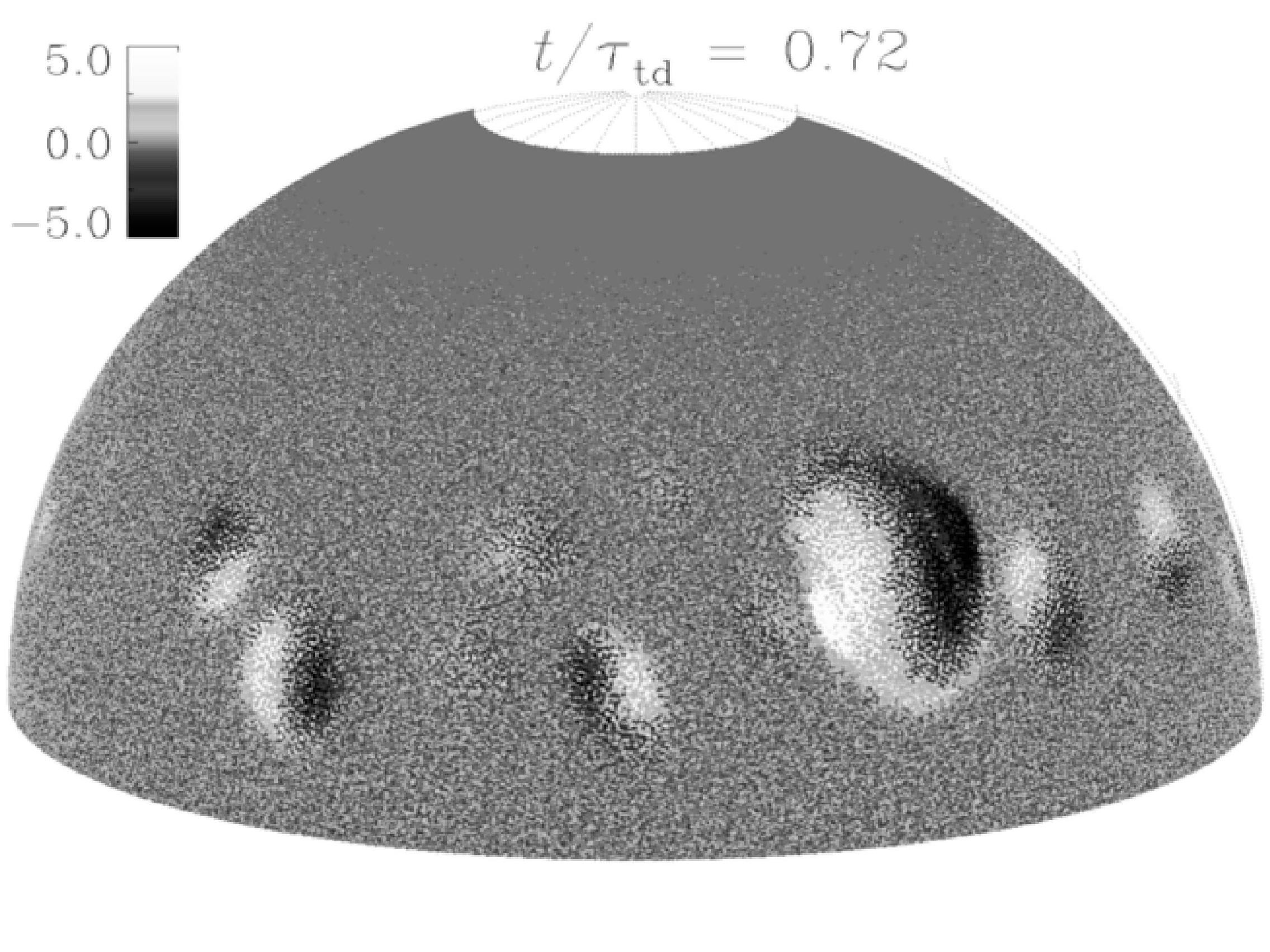}
\caption{
$B_r/\Beq$ \blue{at $r/R=0.98$} for simulations with different stratifications
for Runs~Q1, Q2 and Q3 with density contrasts
2, 450 and 1400 from left to the right, respectively.
%\url{256dyn_sph_sin_g1_vbc_pi}
%\url{256dyn_sph_sin_g14_vbc_pi}
%\url{256dyn_sph_sin_g17_vbc_pi}
}\label{fig:pstr}
\end{figure*}

The time dependence of magnetic energy and its radial profile
are shown in \Fig{fig:keme} \blue{for different normalizations}.
Different colors correspond to different degree of stratification.
Earlier studies have demonstrated the importance of stratification
in the formation of structures through NEMPI.
In fact, by increasing the stratification, the structures were found to
be more intense.
For the highest stratification ($\Gamma_\rho=1400$), the total
magnetic energy becomes somewhat larger than for $\Gamma_\rho=450$,
although the strength of the spots is about the same.
This suggests that, there might be an upper limit for the density contrast.
A similar effect was also observed in the work of \cite{Jab14},
which they referred to it as gravitational quenching, which
saturates or even suppresses NEMPI.
We must also remember that NEMPI can only work in regions where the
magnetic field relative to the equipartition value is in the optimal
range \citep{KBKR12,LBKR14,BGJKR14}.
However, in our strongly stratified system, the regions where this
would be the case can become rather shallow.
This gives another geometric constraint on the possibility of NEMPI,
which was already discussed by \cite{LBKR14} in connection with polytropic
stratifications, where this limitation can become particularly severe.
Further studies are needed to understand the implications of this
geometric effect.

\blue{
The middle panel of \Fig{fig:keme} shows that, in the deeper parts
($r/R\leq0.77$), the normalized magnetic field strength is virtually
independent of stratification.
This demonstrates that the dynamo is not affected by stratification
or the resulting spot formation.
In the upper layers, on the other hand, the ratio $\bra{\BB^2}/\Beq^2(r)$
varies significantly between the runs with different stratification.
This is mainly a consequence of the strong variation of $\Beq$, which
can be seen from the bottom panel where $\bra{\BB^2}/\Beqz^2$ is shown.
}

\blue{
\Fig{fig:xz_slice} shows the time evolution of $B_r/\Beqz$
in the $r\phi$ plane along the active region belt ($\theta=22\degr$).
One can see the formation of dynamo-generated large-scale field
at the bottom and an intense magnetic field concentration at the surface layer.
There is a similarity between \Fig{fig:xz_slice} and Figure~8 of \cite{Mit14},
because in both cases there appears to be a mechanism that concentrates
the dynamo-generated sub-equipartition field in the deeper parts into
super-equipartition field in the upper.
}

\subsection{Inclination angle}

Our spots show a systematic East-West orientation with negative
vertical field on the left and positive values on the right.
In addition, some of the regions also show a certain tilt, although the
apparent yin-yang structure makes it hard to say whether the tilt angle
is positive or negative.
Most of the bipolar regions are oriented in a similar fashion, although
there is also a large fraction of spots that show random orientation.
\cite{Par55} suggested that sunspot pairs are produced by the buoyant
rise of a flux tube, which takes the form of an $\Omega$ loop near the
surface.
To see whether this is also the case in the present simulations, we show
$B_r/\Beq$ at the surface together with field vectors
projected onto the horizontal plane (\Fig{fig:yz_slice30}).
Note that the vectors tend to point in the negative $\phi$ direction,
i.e., the azimuthal field points to the left.
In most of the bipolar spots in this figure, $B_r/\Beq$ tends to
be positive on the right-hand side of the spot (pointing upward) and
negative on the left-hand side of the spot (pointing downward).
This corresponds to the expected $\Omega$ loop scenario.

\begin{figure*}[t!]\begin{center}
\includegraphics[width=.32\textwidth]{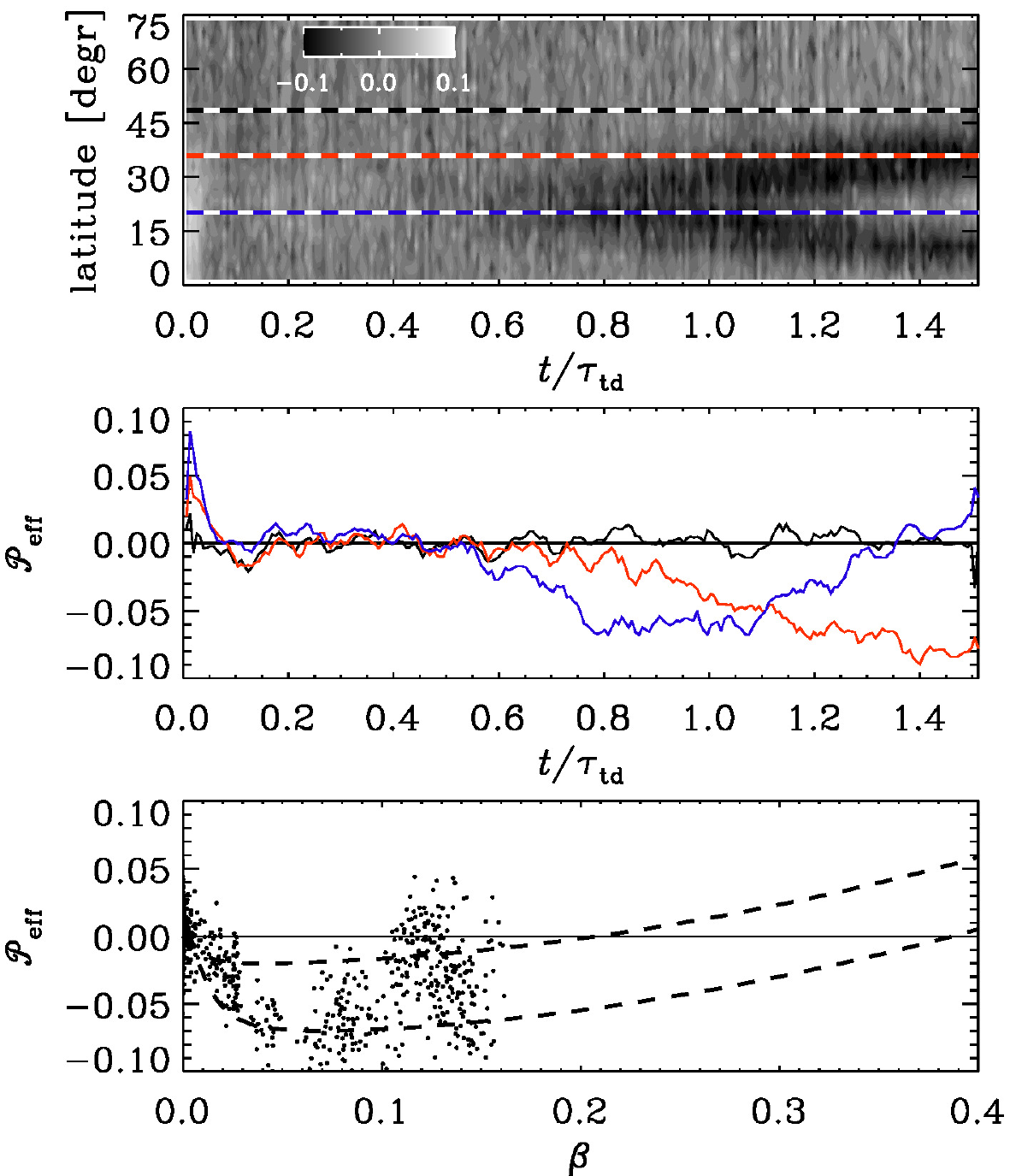}
\includegraphics[width=.32\textwidth]{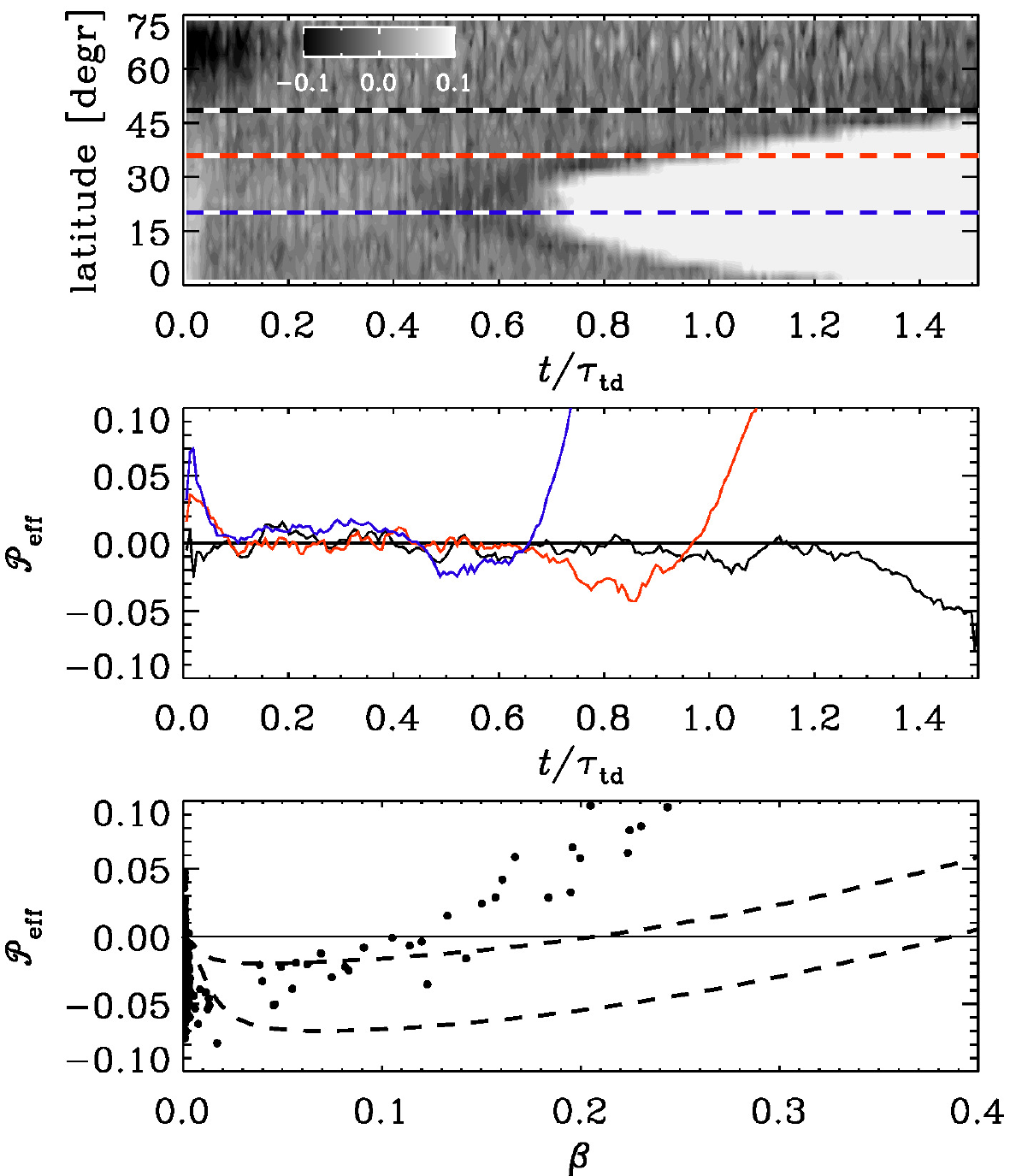}
\includegraphics[width=.32\textwidth]{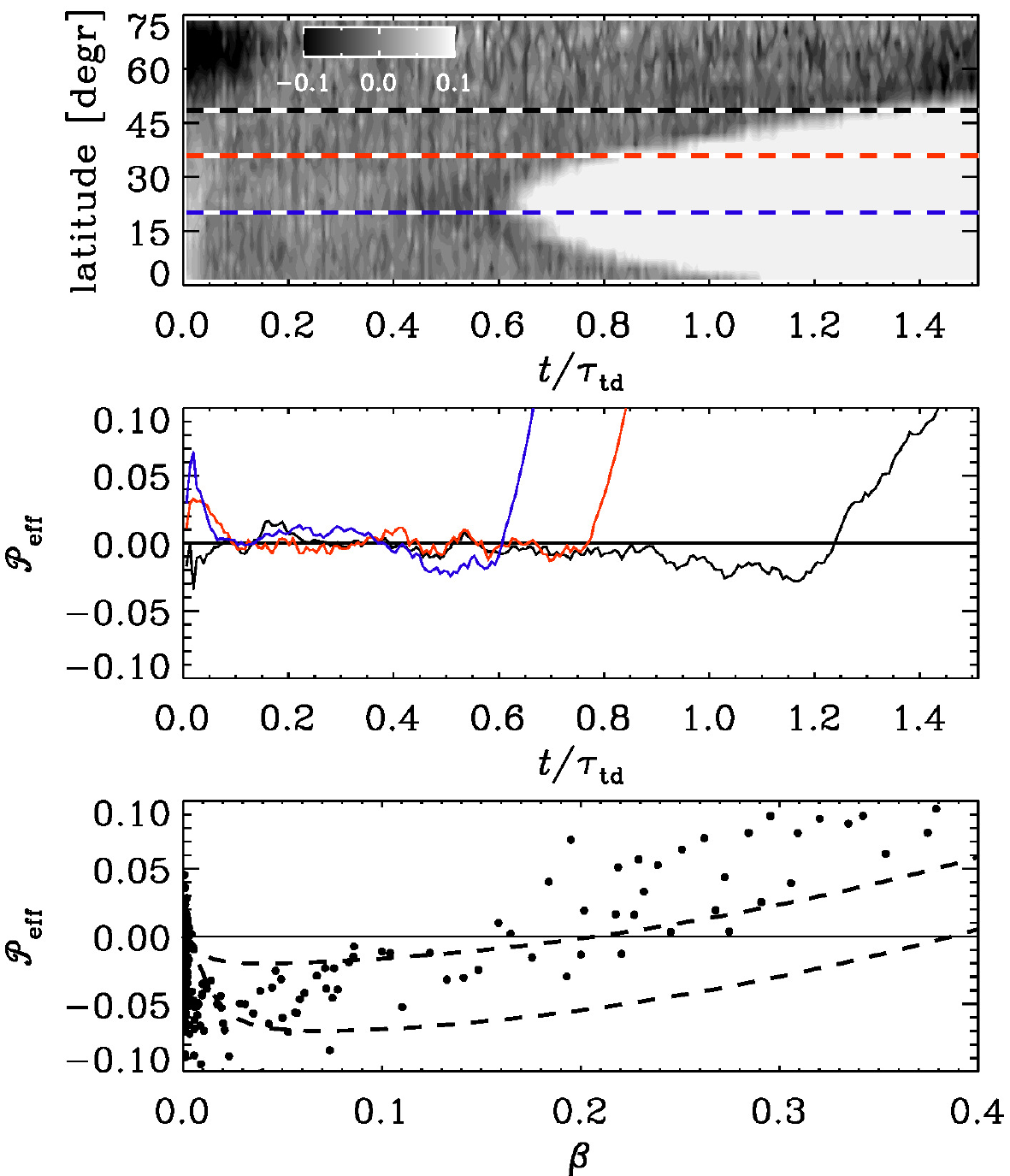}
\end{center}\caption[]{
$\Peff$ vs.\ $t$ and $\theta$ for $r/R=0.85$ in the top panels,
for $\Gamma_\rho=2$ (Run~D1, left), 450 (Run~D2, middle column), and 1400 (Run~D3, right).
The horizontal lines in black, red, and blue indicate three latitudes
(50\degr, 35\degr, and 20\degr), for which we show $\Peff$ vs.\ $t$
in the middle panels.
The bottom panels show scatter plots of $\Peff(\beta)$.
}\label{ppppresst_reduce}\end{figure*}

%\begin{figure}[t!]\begin{center}
%AB: now 3 rows
\begin{figure*}[t!]\begin{center}
%\includegraphics[width=\columnwidth]{ppppresst_reduce_ir2_256dyn_sph_sin_g14_dip_pi_stress}
%AB: now added Sarah's two new figures
\includegraphics[width=.32\textwidth]{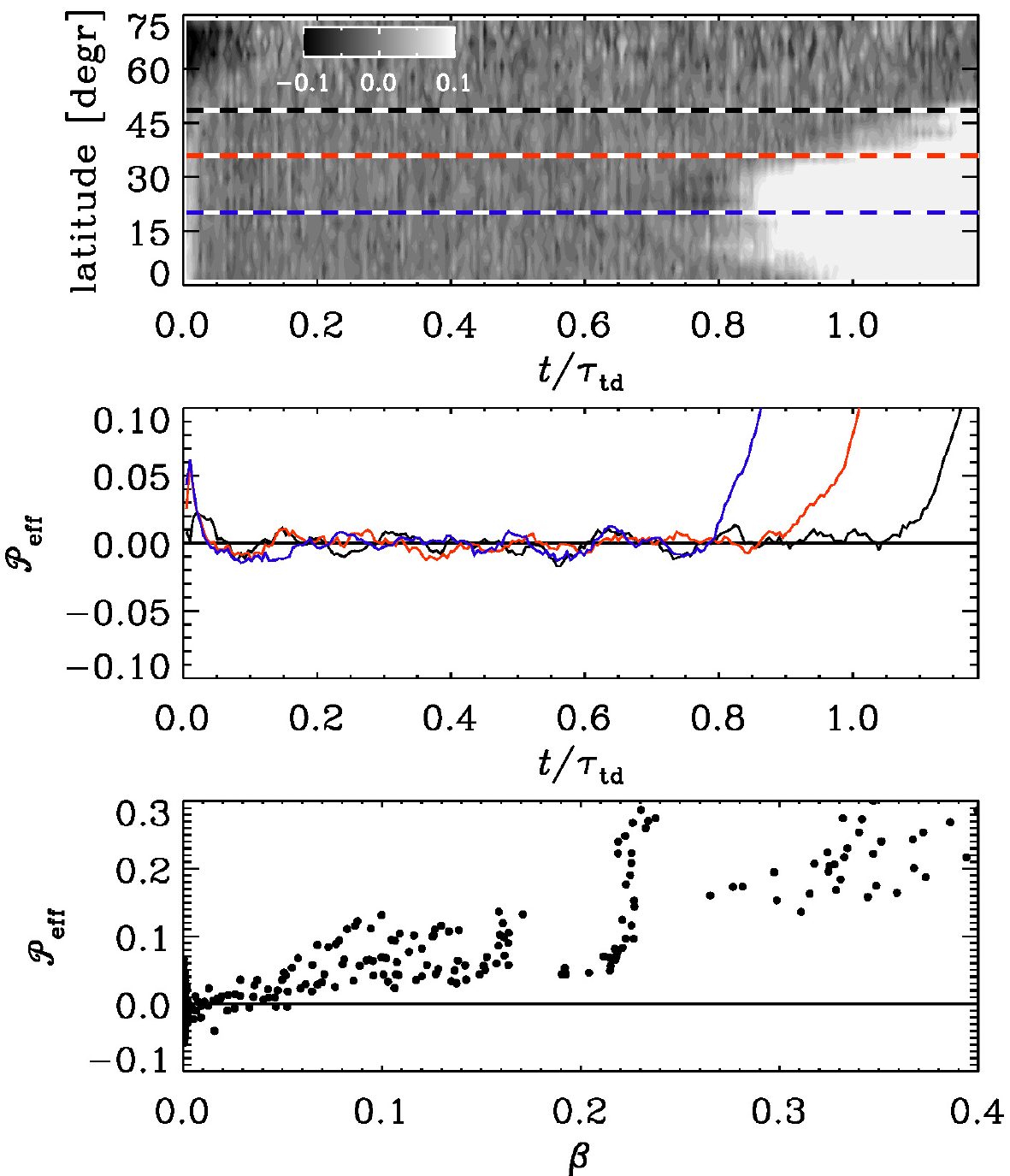}
\includegraphics[width=.32\textwidth]{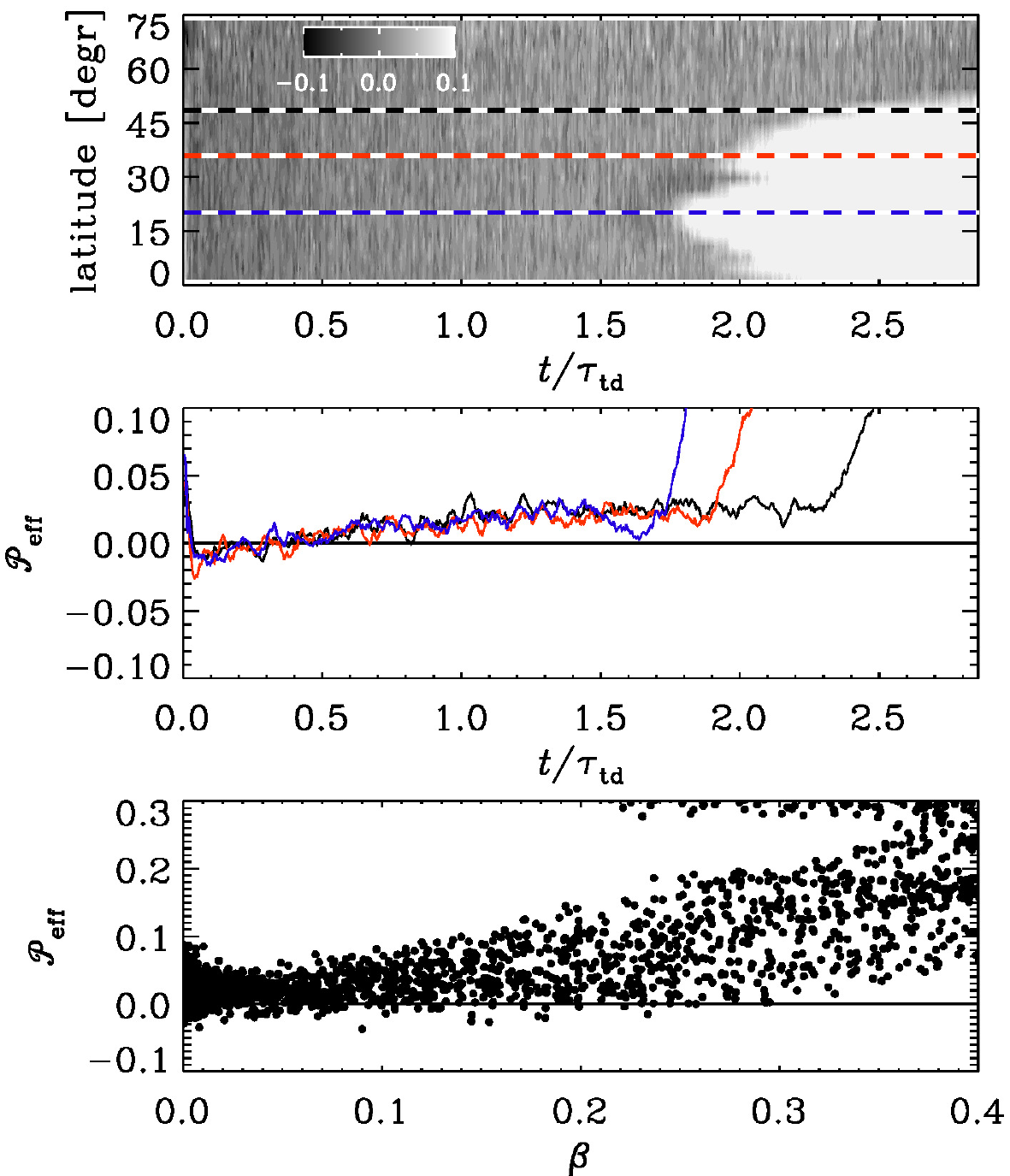}
\includegraphics[width=.32\textwidth]{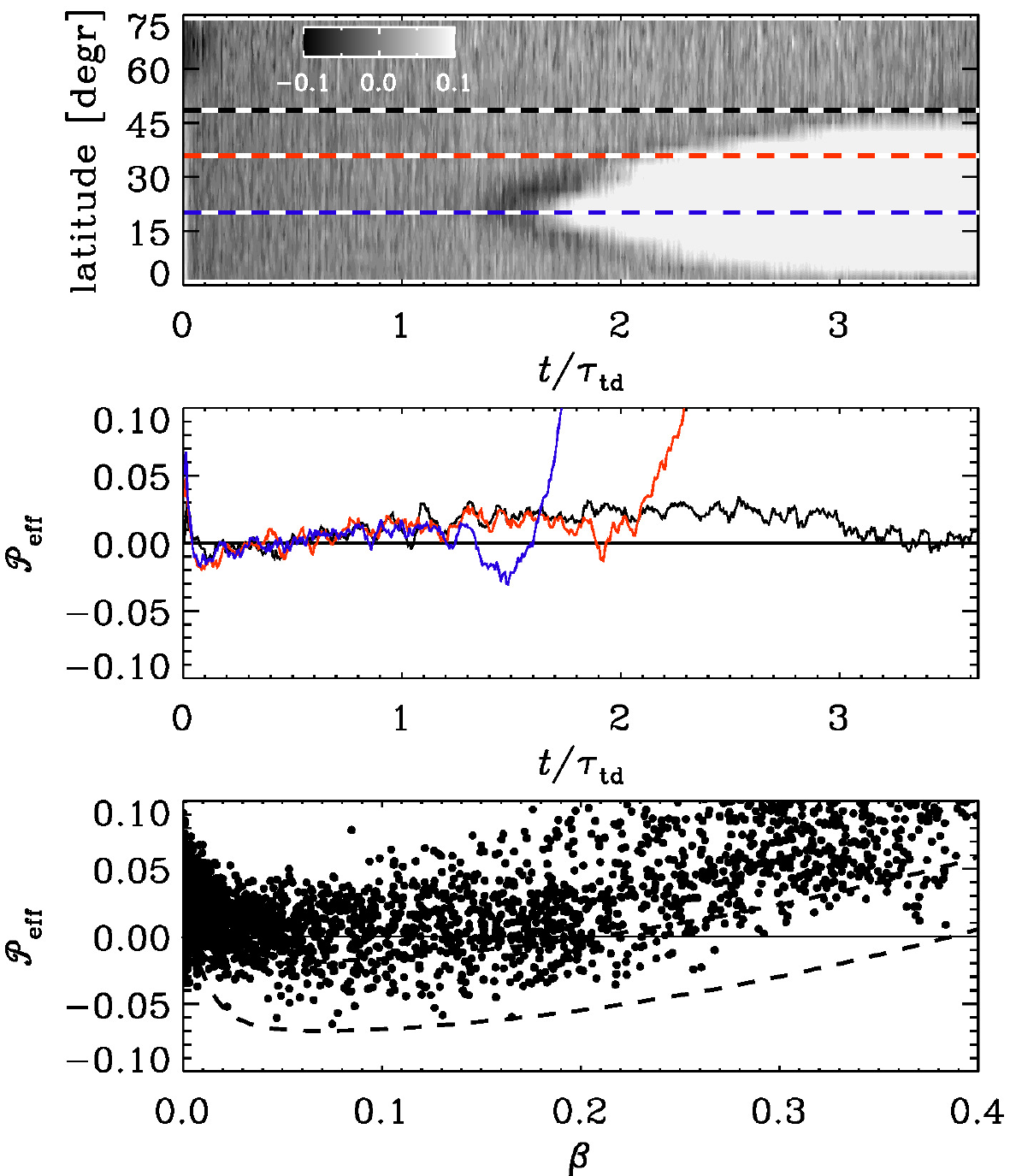}
\end{center}\caption[]{
%Similar to \Fig{ppppresst_reduce}, but for $r/R=0.98$ and for
%AB: now explained Sarah's two new figures (not discussed yet in the text)
Similar to \Fig{ppppresst_reduce}, but for $r/R=0.98$ and for
%$\Gamma_\rho=450$.
%AB: 
$\Gamma_\rho=450$ at $\smax=1$ (left), $\smax=0.5$ (middle), and
$r/R=0.85$, again with $\smax=0.5$ (right).
%}\label{ppppresst_reduce_ir2}\end{figure}
%AB: now 3 rows
}\label{ppppresst_reduce_ir2}\end{figure*}

\begin{figure*}[t!]\begin{center}
\includegraphics[width=.69\columnwidth]{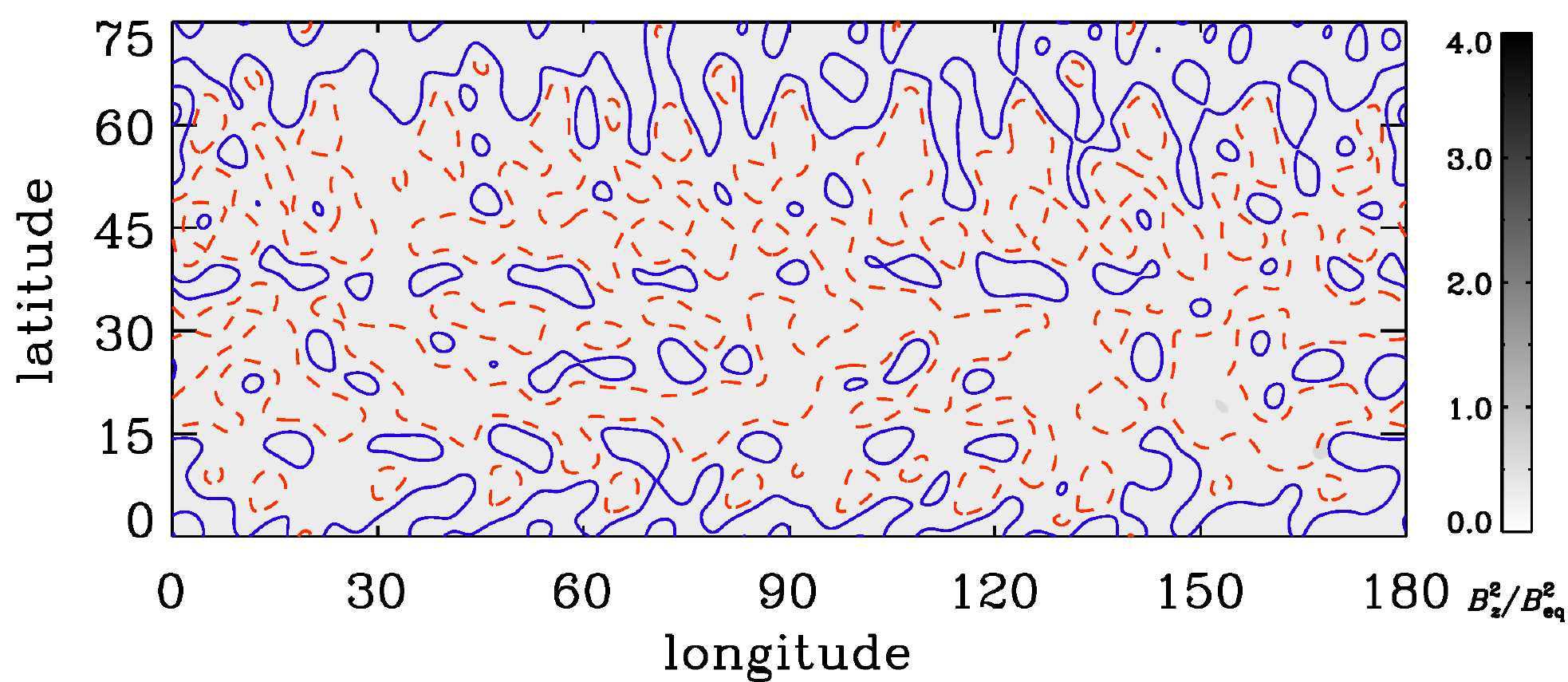}
\includegraphics[width=.69\columnwidth]{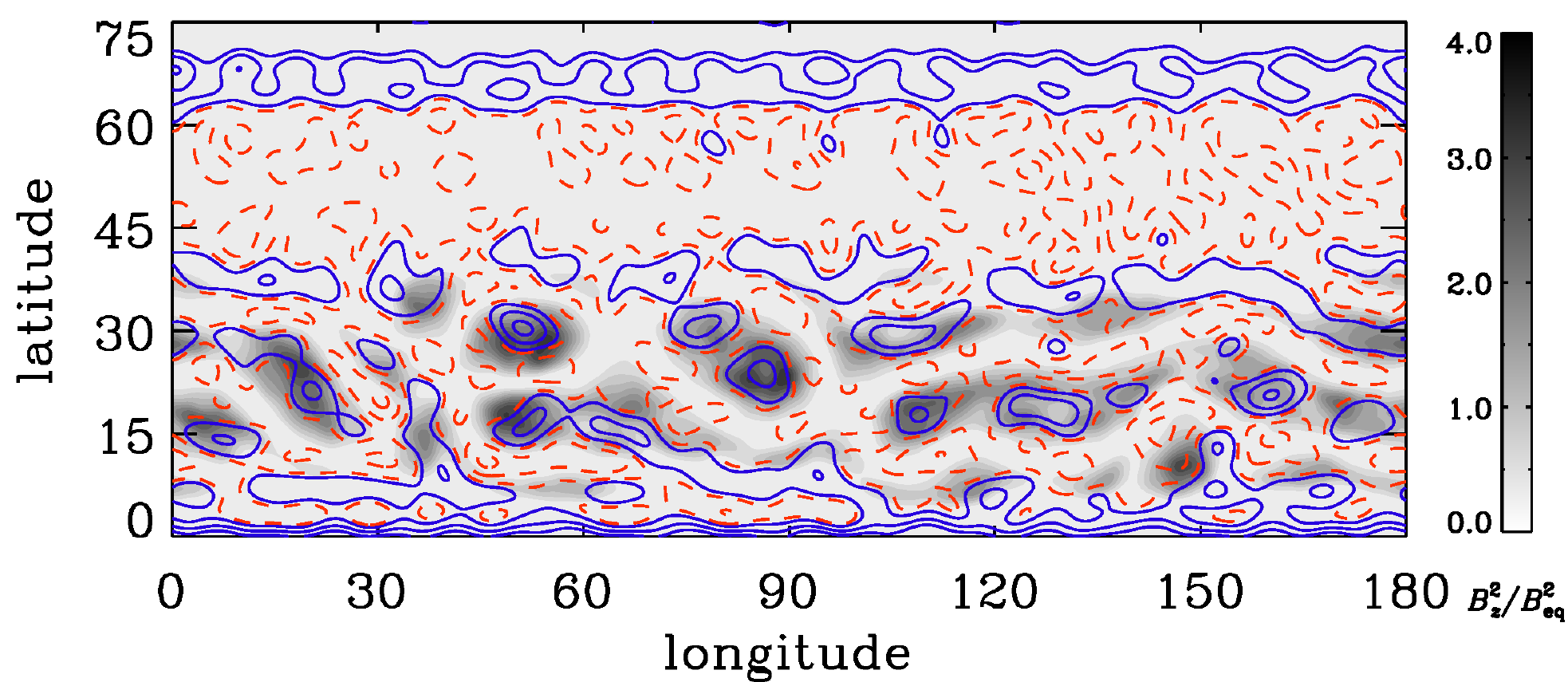}
\includegraphics[width=.69\columnwidth]{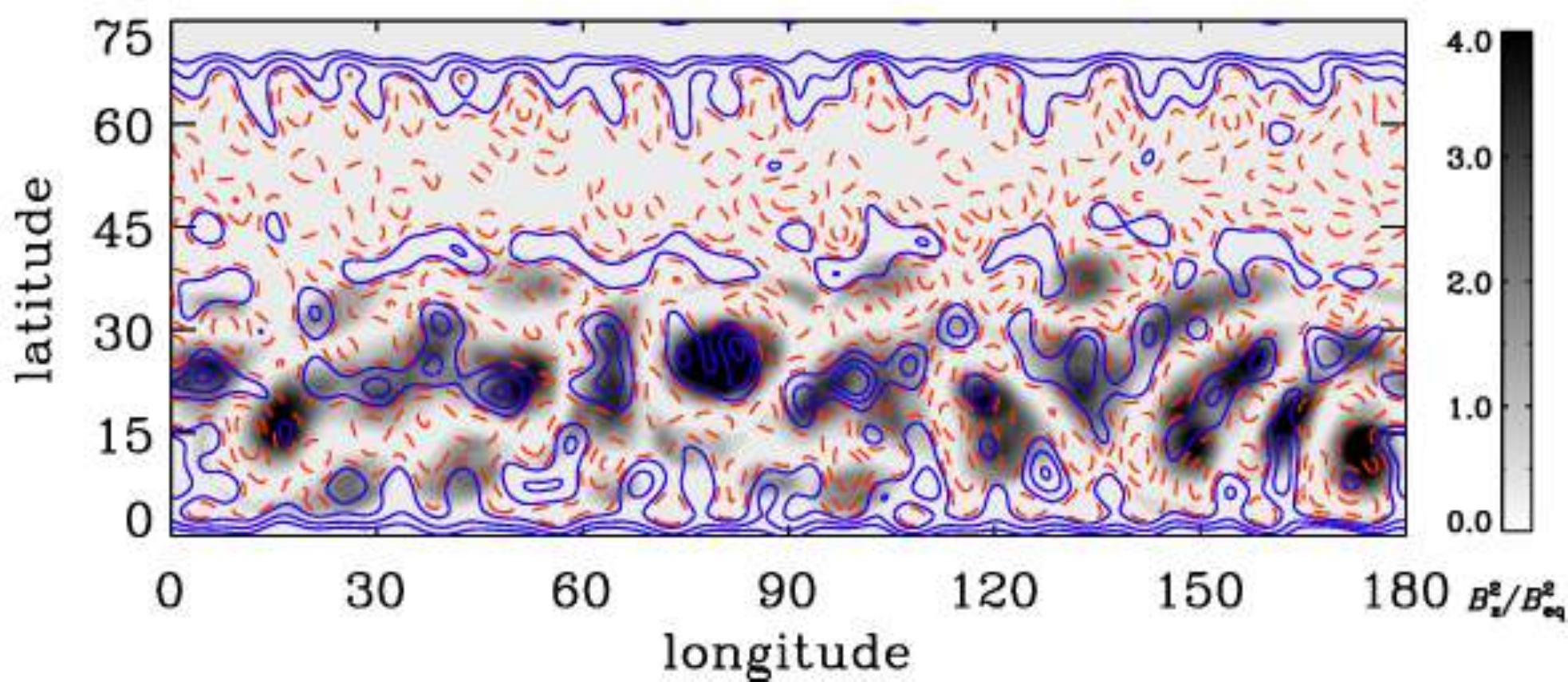}
\end{center}
\caption{
Contours of negative (blue, solid lines) and positive (red, dashed)
vertical velocity $\bra{U_r}_{kR<50}$ superimposed on a gray-scale
representation of $\bra{B_r}_{kR<100}^2/\Beq^2(r)$ in
Mercator projection at $r/R=0.85$ and $t/\tautd=0.7$
\blue{for Run~D1 (left panel, Run~D2 (middle panel) and, Run~D3 (right panel))}.
}\label{fig:pbx_one}
\end{figure*}

%\subsection{Dependence on $\rstar$, $\sigma$, Reynolds number,
%AB: should be smax
\subsection{Dependence on $\rstar$, $\smax$, Reynolds number,
stratification, and scale separation}

We recall that we adopt a similar forcing setup as \cite{Mit14} with
a transition at a radius $\rstar$ from helical forcing in the deeper parts to
non-helical in the upper parts.
\cite{Mit14} found that, when the border is moved closer to the bottom
of the convection zone, the structures appear later.
This is due to the fact that it takes a
longer time for dynamo to affect the upper layers.
A similar behavior was observed when the helicity parameter
%$\sigma$ is decreased from its maximum value of 1.
%AB: should be smax
$\smax$ is decreased from its maximum value of 1.
In such a case the formation of structures occurs again
with time delay, which is due to the weaker dynamo.
\blue{
This is shown in \Fig{fig:psig}, where we present the time evolution of
%$B_r/\Beq$ at $r/R=0.98$ for Run~D5 with $\sigma=0.5$.
%AB: should be smax
$B_r/\Beq$ at $r/R=0.98$ for Run~D5 with $\smax=0.5$.
One can see the appearance of spots at later times ($t/\tautd\approx2$).
\Fig{fig:psig} also illustrates the fact that by weakening the helicity
the spots become more intense and regular in comparison with the fully
helical case (compare with \Fig{fig:pptQ}).}
However, the present simulations show that the dynamo
growth rate does even slightly increase with decreasing
helicity; cf.\ Runs~D2 and D4 \blue{or D5}.
On the other hand, lowering the value of $\rstar$ does lead to a small
decrease of the growth rate; cf.\ Runs~Q2 and Q4 as well as Runs~H2 and H4.

The effect of stratification on the formation of the spot is shown
in \Fig{fig:pstr}.
For the lowest density contrast, no structures form, independently of time,
size of the shell, value of Reynolds number, and position of the border
between helical and non-helical turbulence.
This confirms that stratification plays a crucial role in the formation of
bipolar regions.
\blue{
Simulations with different density contrasts ($\Gamma_\rho$ between 2 and 1450)
show that bipolar structures form for $\Gamma_\rho$ larger than 30.
We note that for weak stratification ($\Gamma_\rho=30$ for instance)
spots appear only at late times ($t/\tautd\approx3$), while
for stronger stratifications ($\Gamma_\rho=70$) they appear earlier
($t/\tautd\approx1$) and are more concentrated.}
\blue{
In the deeper parts ($r/R=0.75$), on the other hand, the magnetic field
evolution for Run~Q1 with weak stratification is similar to that of Run~Q2,
which demonstrates that the dynamo is not effected by the strength of
stratification.
One also sees that the late time evolution of the magnetic field
is similar for Runs~Q1 and Q2, even though spots form in Run~Q2,
but not in Q1.
This suggests that the formation of spots does not affect the dynamo.
}
In spherical geometry, it is possible to investigate the effect of
different $\phi$ extent on the formation of structures.
Not surprisingly, it turns out that for small $\phi$ extent (below $\pi/6$)
the size of the spots is limited by the domain size, because the formation of
strongly inhomogeneous structures requires strong scale separation 
between the energy-carrying eddies and the domain.

\subsection{Effective magnetic pressure}

To assess whether NEMPI is operating, we now
calculate the effective magnetic pressure, following \cite{BKKR12}
and adapting the formulation to the spherical case.
\blue{
The total stress from the fluctuating velocity and magnetic fields
is given by
\EQ
\overline\Pi_{ij}^{\rm f}
=\meanrho\,\overline{u_i u_j}
+\half\delta_{ij}\overline{\bb^2}
-\overline{b_i b_j},
\label{PPij}
\EN
where the superscript $f$ denotes the fluctuating terms.
In the following we only need the three diagonal components
of $\overline\Pi_{ij}^{\rm f}$, which we denote by
$\overline\Pi_{i}^{\rm f}$, where $i$ refers to
\blue{$r$,} $\theta$, or $\phi$.
As we are interested in the contribution from the part
that results from the mean field, we should calculate the stress also
for zero mean field and subtract it from the total stress, so}
\EQ
\Delta\overline\Pi_{i}^{\rm f}
=\meanrho\,(\overline{u_i^2}-\overline{u_{0i}^2})
+\half(\overline{\bb^2}-\overline{\bb_0^2})
-(\overline{b_i^2}-\overline{b_{0i}^2}),
\label{Pi_ii}
\EN
where subscript $0$ refers to the case with zero mean magnetic field.
But as the background field is here dynamo-generated,
we use the values of the related quantities in the upper part of the domain
(non-helical part) at early times.
This gives us the possibility to estimate the effective magnetic pressure in
spherical geometry with dynamo-generated magnetic field.
In \Eq{Pi_ii}, $\Delta\overline\Pi_{i}^{\rm f}$ denotes the three
diagonal components of the tensor $\Delta\overline\Pi_{ij}^{\rm f}$.
In the mean-field description, it depends on the
mean magnetic field $\meanBB$ and is parameterized as
\begin{eqnarray}
\Delta\overline\Pi_{ij}^{\rm f}
= \left(q_{\rm s} \hat \beta_i \hat \beta_j
- \half q_{\rm p} \, \delta_{ij} + q_{\rm g} \,
\ghat_i \, \ghat_j \right)
\meanBB^2 ,
\label{funct-tenzor}
\end{eqnarray}
where $\hat \beta_i$ and $\ghat_i$ are the unit vectors along $\meanBB$
and ${\bm g}$, respectively.
The effective magnetic pressure is defined as a sum of
non-turbulent and turbulent contributions:
\EQ
\Peff=\half(1-\qp)\beta^2,
\EN
where $\beta^2=\overline{B}^2/B_{\rm eq}^2$ and $\qp$ is a 
turbulent transport coefficient that depends on 
the mean magnetic field and can be computed from
the DNS as
\blue{
\begin{eqnarray}
q_{\rm p}=-\frac{1}{\meanBB^2}[\Delta \overline\Pi_{\theta}^{\rm f}+\Delta \overline\Pi_{\phi}^{\rm f}
-q_{\rm s}(\meanB_\theta^2+\meanB_\phi^2)],\label{qp}
\end{eqnarray}
with
\begin{eqnarray}
q_{\rm s}=\left.\left(\Delta \overline\Pi_{\theta}^{\rm f}-\Delta \overline\Pi_{\phi}^{\rm f}\right)\right/
\left(\meanB_\theta^2-\meanB_\phi^2\right),
\label{qs}
\end{eqnarray}
and
\begin{eqnarray}
q_{\rm g}=\frac{1}{\meanBB^2}\left[-\Delta \overline\Pi_{r}^{\rm f}+q_{\rm s}\meanB_r^2-
\frac{1}{2}q_{\rm p}\meanBB^2\right].
\label{qg}
\end{eqnarray}
}

Previous studies \citep{BKKR12,KBKMR12} in Cartesian geometry have shown
that $q_{\rm s}$ and $q_{\rm g}$
are very close to zero.
This is also confirmed by the present simulations, where
\blue{$\Delta \overline\Pi_{\theta}^{\rm f}-\Delta \overline\Pi_{\phi}^{\rm f}$}
is found to correlate poorly with $\meanB_\theta^2-\meanB_\phi^2$.
We therefore ignore $q_{\rm s}$ and $q_{\rm g}$ in most of the following.

Next, we subtract the time average of an early time interval
between times $t_1$  and $t_2$ and compute the diagonal components
of the change of the stress as
\blue{
\begin{equation}
\Delta\overline{\Pi}_i^{\rm f}(r,\theta,t)=\overline{\Pi}_i^{\rm f}(r,\theta,t)
-{1\over t_2-t_1}\int_{t_1}^{t_2}\overline{\Pi}_i^{\rm f}(r,\theta,t')\,\dd t',
\end{equation}
where $t_1=0.1\tautd$ and $t_2=0.5\tautd$ denote the time interval over
which the turbulence in the upper layer
is not yet affected by the mean magnetic field.
}
We thus compute
\blue{
\begin{equation}
\qp=-\left.\left(\Delta\overline{\Pi}_\theta^{\rm f}
+\Delta\overline{\Pi}_\phi^{\rm f}\right)\right/\meanB^2,
\end{equation}
}
so we get $\Peff=\half(1-\qp)\beta^2$.

In \Fig{ppppresst_reduce} we show $\Peff$ vs.\ $t$ and $\theta$ for $r/R=0.85$
using Runs~Q1, Q2, and Q3.
The bottom panels show scatter plots of $\Peff(\beta)$.
It is customary to fit such data to an expression of the form \citep{KBKR12}
\begin{equation}
\qp(\beta)=\beta_\ast^2/(\betap^2+\beta^2),
\label{qpfit}
\end{equation}
where $(\beta_\ast,\betap)$ is a set of fit parameters.
They have previously been determined for Cartesian simulations with an
imposed magnetic field \citep{BKKR12}.
In \Fig{ppppresst_reduce}, the dashed lines show for comparison
the result for two representations of \Eq{qpfit}
with $(\beta_\ast,\betap)=(0.39,0.013)$ and $(0.21,0.008)$
for curves with the deeper and less deep minimum, respectively.
Neither of the curves fit the data points well.
Nevertheless, it is important to note that $\Peff$ is always negative
for $\beta<0.1$, although in the runs with stronger stratification
the number of such points is rather small.

\blue{
One may speculate that in Runs~D2 and D3, which produce strong fields,
there are two stages of magnetic field concentrations.
At early times, the field is below the equipartition field strength,
so NEMPI works and the effective magnetic pressure is negative.
At later times, when the field is of the order of or larger than the
equipartition field strength, the effective magnetic pressure becomes
positive and the standard magnetic buoyancy instability might play a
role in bringing magnetic field to the surface, perhaps as what is seen
in \Fig{fig:xz_slice}.
We may conclude that, while there is evidence for
negative values of $\Peff$, there may still be other effects
playing important roles.
In particular the formation of relatively sharp boundaries of our spots,
which is a marked feature of both the present calculations and those of
\cite{Mit14}, may be due to such a new effect.
It is reminiscent of the appearance of sharp structures as a result of
ambipolar diffusion under laminar conditions \citep{BZ94}.
If so, it may be related to mean-field terms in the induction equation
rather than the momentum equation.
}

In the near-surface layers, on the other hand, there is no evidence
for negative effective magnetic pressure, as can be seen from
%\Fig{ppppresst_reduce_ir2}, where we show the results for $\Peff$
%AB: mention left panel
\Fig{ppppresst_reduce_ir2} (left panels), where we show the results for $\Peff$
at $r/R=0.98$ and for $\Gamma_\rho=450$.
This is however consistent with the idea that spots are formed
mainly due to suction from deeper layers \citep{BGJKR14}.
\blue{
%AB: added this
For Run~D5 with $\smax=0.5$, the results are similar to Run~D2,
although $\Peff$ is slightly smaller (middle panels of
\Fig{ppppresst_reduce_ir2}) and at $r/R=0.85$ the values of $\Peff$ are
generally less negative (right panels of \Fig{ppppresst_reduce_ir2}).
%AB.
}

To check whether there are downflows at greater depth where NEMPI can act,
we compare in \Fig{fig:pbx_one} for \blue{Runs~D1,} D2 \blue{, and, D3}
contours of negative and positive
vertical velocity, $\bra{U_r}_{kR<50}$, superimposed on a gray-scale
representation of $\bra{B_r}_{kR<100}^2/\Beq^2(r)$ at $r/R=0.85$.
Here, $\bra{\cdot}_{kR}$ denotes Fourier filtering, applied to obtain
smoother contours.
(For $B_r$ we also apply some filtering, but only above $kR=100$
to eliminate patterns on the scale of the forcing.)
We see that, \blue{for sufficiently strong stratification} (Runs~D2 and D3),
there are indeed many locations where the field is strong and
$\bra{U_r}_{kR<50}$ is negative, but the correlation is not very strong.
However, the speed of the downflows increases with increasing
stratification (compare the last two panels of \Fig{fig:pbx_one}).
Furthermore, since the field strength exceeds the equipartition value
even at this greater depth, NEMPI must have seized to work.
Nevertheless, it could have operated at earlier times when the field
was weaker.

Finally, a comment regarding the possible importance of the $\qs$ term is
in order.
As we have mentioned above, $\qs$ cannot be determined owing to the poor
correlation between $\Delta \overline\Pi_{\theta}-\Delta \overline\Pi_{\phi}$
and $\meanB_\theta^2-\meanB_\phi^2$.
However, at $r/R=0.85$ and near $35\degr$ latitude, the correlation
is not quite as poor and $\qs$ can be determined in the time interval
$0.7<t/\tautd<0.85$, where for Run~D2 $\qs\approx20$ is found.
On the other hand, in the neighborhood of this latitude, $\qs$ turns out
to be in the range $0<\qs<20$.
The corresponding values of $\qg$ are found to be in the range
from $-5$ to $10$.
Furthermore, if the $\qs$ term in \Eq{qg} were ignored, we find values
in the range $-5<\qg<0$.
Interestingly, mean-field simulations of NEMPI have shown earlier that
values in the range $-10<\qg<10$ do not significantly affect the growth rate
\citep{KBKMR12}, suggesting that $\qg$ remains subdominant, regardless of
whether or not the $\qs$ term is taken into account.

\begin{figure}[t]
\centering
\includegraphics[width=0.7\columnwidth]{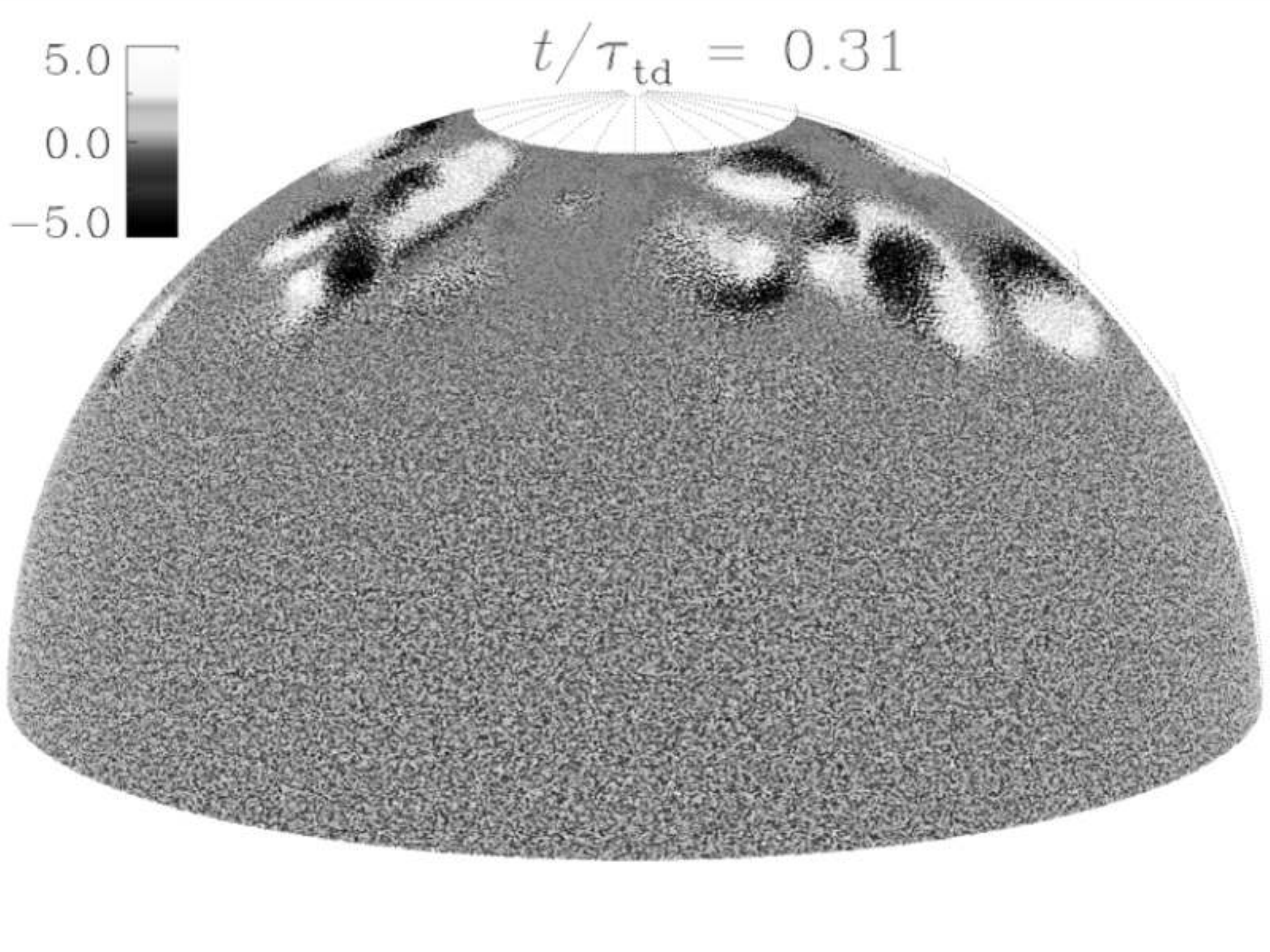}
\caption{
Formation of the high-latitude spots for the case $n=0$ (Run H2).
}\label{fig:highlatitude}
\end{figure}

\subsection{High-latitude spots}
\label{polar spot}

Before concluding, let us return once more to the occurrence of magnetic
spots and its dependence on the parameter $n$ in \Eq{alpha_profile}.
When the helicity is large at high latitudes,
i.e., when we choose the helicity profile with $n=0$,
magnetic spots are found to form close to the poles in a fashion reminiscent
of recent simulations by \cite{Yadav}.
\Fig{fig:highlatitude} presents the formation of \blue{bipolar} spots
near the pole for run H2.
\blue{
By contrast, the work of \cite{Yadav} showed just a single spot.
However, in both cases the underlying dynamo process is a distributed
one, so the lower boundary at the bottom of the domain is not critical
for its operation.
}

The results of the simulations with $n=0$ show similar behavior
and parameter dependence as the case with $n=6$.
As for $n=6$, for weaker stratification no spots form and for smaller
%$\sigma$ and deeper $\rstar$, the structures form with a time delay.
%AB: should be smax
$\smax$ and deeper $\rstar$, the structures form with a time delay.

\section{Conclusions}

The present work has demonstrated that in a strongly stratified two-layer
spherical model with helical turbulence
in the lower layer and non-helical turbulence in the upper one,
the $\alpha^2$ dynamo produces large-scale magnetic fields that develop
sharp spot-like structures at the surface.
This extends the results of \cite{Mit14} to spherical geometries.
We therefore see for the first time that
the bipolar magnetic spots have a finite size \blue{that is
not limited by the domain size as in the work of \cite{Mit14}.}
However, contrary to earlier expectations \citep{KBKR12,BGJKR14}, the size
of these structures exceeds the local value of the density scale height
by much more than the earlier expected value of about ten.

In our present simulations the dynamo (caused by the $\alpha^2$ dynamo)
\blue{
was very efficient, because the forcing was assumed to be fully helical.
In reality, the helicity is caused by the combined action of rotation
and stratification \citep{KR80}.
In this sense, dynamos do depend on stratification, contrary to the
present case with helical forcing, where this was found to be not the
case; see \Sec{SpotFormation}.
As a consequence of the strong helicity,}
the resulting large-scale field is rather strong and
the magnetic spots begin to fill eventually the entire horizontal surface.
This is also what is expected for very active stars where the filling
factor of the surface magnetic field is known to reach unity as the star
becomes more active \citep{SL85}.
Conversely, to model sunspots, which are much smaller, we expect
that we would need to decrease the fractional helicity below the
values explored in the present work.

\blue{
Visualizations of the three-dimensional magnetic field structure in
convectively driven dynamos has revealed the formation and subsequent
rise of serpentine-shaped flux tubes \citep{NM14,FF14}.
Our work has now shown that strong stratification may provide the key to
understanding how such structures can experience re-amplification, which
is required if they are to be responsible for spot formation.
}

While it is clear that strong stratification
and large scale separation between the turbulent integral scale
and the size of the box are essential elements
behind spot formation, the results presented here are different from our
earlier findings that were obtained under more idealized conditions
such as the use of an imposed magnetic field.
However, like the earlier results of \cite{Mit14}
we find again evidence for downflows below the sites of spot formation
and, in particular, the formation of sharply bounded structures.
The latter is reminiscent of the appearance of sharp structures as a result of
ambipolar diffusion under laminar conditions \citep{BZ94}.
This raises the question, whether for strongly stratified turbulence
the effective magnetic diffusivity can attain a nonlinear dependence
on the magnetic field that is similar to that of ambipolar diffusion,
i.e., it increases with increasing field strength.
To address these questions, it would be best to return to Cartesian
geometry where it would be possible to determine turbulent transport
coefficients with dedicated methods such as the test-field approach.
This is however beyond the scope of the present work.

\acknowledgments
This work was supported in part by
the Swedish Research Council Grants No.\ 621-2011-5076 and 2012-5797,
the Research Council of Norway under the FRINATEK grant 231444,
the Academy of Finland under the ABBA grant No.~280700, as well as
the Russian Federal Program under the grant No.~14.578.21.0033.
We acknowledge the allocation of computing resources provided by the
Swedish National Allocations Committee at the Center for
Parallel Computers at the Royal Institute of Technology in
Stockholm and the National Supercomputer Centers in Link\"oping,
the High Performance Computing Center North in Ume\aa,
and the Nordic High Performance Computing Center in Reykjavik.

%  journals
\newcommand{\yastroph}[2]{ #1, astro-ph/#2}
\newcommand{\ycsf}[3]{ #1, {Chaos, Solitons \& Fractals,} {#2}, #3}
\newcommand{\yepl}[3]{ #1, {Europhys.\ Lett.,} {#2}, #3}
\newcommand{\yaj}[3]{ #1, {AJ,} {#2}, #3}
\newcommand{\yjgr}[3]{ #1, {J.\ Geophys.\ Res.,} {#2}, #3}
\newcommand{\ysol}[3]{ #1, {Sol.\ Phys.,} {#2}, #3}
\newcommand{\yapj}[3]{ #1, {ApJ,} {#2}, #3}
\newcommand{\ypasp}[3]{ #1, {PASP,} {#2}, #3}
\newcommand{\yapjl}[3]{ #1, {ApJ,} {#2}, #3}
\newcommand{\yapjs}[3]{ #1, {ApJS,} {#2}, #3}
\newcommand{\yija}[3]{ #1, {Int.\ J.\ Astrobiol.,} {#2}, #3}
\newcommand{\yan}[3]{ #1, {Astron.\ Nachr.,} {#2}, #3}
\newcommand{\yzfa}[3]{ #1, {Z.\ f.\ Ap.,} {#2}, #3}
\newcommand{\ymhdn}[3]{ #1, {Magnetohydrodyn.} {#2}, #3}
\newcommand{\yana}[3]{ #1, {A\&A,} {#2}, #3}
\newcommand{\yanas}[3]{ #1, {A\&AS,} {#2}, #3}
\newcommand{\yanar}[3]{ #1, {A\&A Rev.,} {#2}, #3}
\newcommand{\yass}[3]{ #1, {Ap\&SS,} {#2}, #3}
\newcommand{\ygafd}[3]{ #1, {Geophys.\ Astrophys.\ Fluid Dyn.,} {#2}, #3}
\newcommand{\ygrl}[3]{ #1, {Geophys.\ Res.\ Lett.,} {#2}, #3}
\newcommand{\ypasj}[3]{ #1, {Publ.\ Astron.\ Soc.\ Japan,} {#2}, #3}
\newcommand{\yjfm}[3]{ #1, {J.\ Fluid Mech.,} {#2}, #3}
\newcommand{\ypepi}[3]{ #1, {Phys.\ Earth Planet.\ Int.,} {#2}, #3}
\newcommand{\ypf}[3]{ #1, {Phys.\ Fluids,} {#2}, #3}
\newcommand{\ypfb}[3]{ #1, {Phys.\ Fluids B,} {#2}, #3}
\newcommand{\ypp}[3]{ #1, {Phys.\ Plasmas,} {#2}, #3}
\newcommand{\ysov}[3]{ #1, {Sov.\ Astron.,} {#2}, #3}
\newcommand{\ysovl}[3]{ #1, {Sov.\ Astron.\ Lett.,} {#2}, #3}
\newcommand{\yjetp}[3]{ #1, {Sov.\ Phys.\ JETP,} {#2}, #3}
\newcommand{\yphy}[3]{ #1, {Physica,} {#2}, #3}
\newcommand{\yaraa}[3]{ #1, {ARA\&A,} {#2}, #3}
\newcommand{\yanf}[3]{ #1, {Ann. Rev. Fluid Mech.,} {#2}, #3}
\newcommand{\yrpp}[3]{ #1, {Rep.\ Prog.\ Phys.,} {#2}, #3}
\newcommand{\yprs}[3]{ #1, {Proc.\ Roy.\ Soc.\ Lond.,} {#2}, #3}
\newcommand{\yprt}[3]{ #1, {Phys.\ Rep.,} {#2}, #3}
\newcommand{\yprl}[3]{ #1, {Phys.\ Rev.\ Lett.,} {#2}, #3}
\newcommand{\yphl}[3]{ #1, {Phys.\ Lett.,} {#2}, #3}
\newcommand{\yptrs}[3]{ #1, {Phil.\ Trans.\ Roy.\ Soc.,} {#2}, #3}
\newcommand{\ymn}[3]{ #1, {MNRAS,} {#2}, #3}
\newcommand{\ynat}[3]{ #1, {Nature,} {#2}, #3}
\newcommand{\yptrsa}[3]{ #1, {Phil. Trans. Roy. Soc. London A,} {#2}, #3}
\newcommand{\ysci}[3]{ #1, {Science,} {#2}, #3}
\newcommand{\ysph}[3]{ #1, {Solar Phys.,} {#2}, #3}
\newcommand{\ypr}[3]{ #1, {Phys.\ Rev.,} {#2}, #3}
\newcommand{\ypre}[3]{ #1, {Phys.\ Rev.\ E,} {#2}, #3}
\newcommand{\ypnas}[3]{ #1, {Proc.\ Nat.\ Acad.\ Sci.,} {#2}, #3}
\newcommand{\yicarus}[3]{ #1, {Icarus,} {#2}, #3}
\newcommand{\yspd}[3]{ #1, {Sov.\ Phys.\ Dokl.,} {#2}, #3}
\newcommand{\yjcp}[3]{ #1, {J.\ Comput.\ Phys.,} {#2}, #3}
\newcommand{\yjour}[4]{ #1, {#2}, {#3}, #4}
\newcommand{\yprep}[2]{ #1, {\sf #2}}
\newcommand{\ybook}[3]{ #1, {#2} (#3)}
\newcommand{\yproc}[5]{ #1, in {#3}, ed.\ #4 (#5), #2}
\newcommand{\pproc}[4]{ #1, in {#2}, ed.\ #3 (#4), (in press)}
\newcommand{\pprocc}[5]{ #1, in {#2}, ed.\ #3 (#4, #5)}
\newcommand{\pmn}[1]{ #1, {MNRAS}, to be published}
\newcommand{\pana}[1]{ #1, {A\&A}, to be published}
\newcommand{\papj}[1]{ #1, {ApJ}, to be published}
\newcommand{\ppapj}[3]{ #1, {ApJ}, {#2}, to be published in the #3 issue}
\newcommand{\sprl}[1]{ #1, {PRL}, submitted}
\newcommand{\sapj}[1]{ #1, {ApJ}, submitted}
\newcommand{\sana}[1]{ #1, {A\&A}, submitted}
\newcommand{\smn}[1]{ #1, {MNRAS}, submitted}

\end{document}